\documentclass[12pt]{article}%
\usepackage{graphicx}
\usepackage{amsmath}
\usepackage{amsfonts}
\usepackage{amssymb}%
\newtheorem{theorem}{Theorem}

\newtheorem{definition}[theorem]{Definition}

\newtheorem{lemma}[theorem]{Lemma}

\newtheorem{proposition}[theorem]{Proposition}
\newtheorem{remark}[theorem]{Remark}

\newenvironment{proof}[1][Proof]{\noindent\textbf{#1.} }{\ \rule{0.5em}{0.5em}}
\begin{document}

\title{\textbf{Dobrushin Interfaces via Reflection Positivity}}
\author{Senya Shlosman\thanks{Also at the Institute for the Information Transmission
Problems, Moscow, Russia, shlos@iitp.ru} and Yvon Vignaud\\Centre de Physique Theorique, UMR 6207 CNRS,\\Luminy Case 907,\\13288 Marseille, Cedex 9, France\\shlosman@cpt.univ-mrs.fr\\vignaud@cpt.univ-mrs.fr}
\maketitle

\begin{abstract}
We study the interfaces separating different phases of 3D systems by means of
the Reflection Positivity method. We treat discrete non-linear sigma-models,
which exhibit power-law decay of correlations at low temperatures, and we
prove the rigidity property of the interface.

Our method is applicable to the Ising and Potts models, where it simplifies
the derivation of some known results. The method also works for large-entropy
systems of continuous spins.

\end{abstract}

\section{Introduction}

The first example of a pure state describing the coexistence of phases
separated by an interface was discovered by R. Dobrushin in 1972, \cite{D72}.
There he was studying the low temperature 3D Ising model. He was considering
the Ising spins in a cubic box $V_{N}$ with $\left(  \pm\right)
$\textit{-boundary condition }$\sigma^{\pm}$: all spins of $\sigma^{\pm}$ are
$\left(  +\right)  $ in the upper half-space and $\left(  -\right)  $ $~$in
the lower half-space. Such a boundary condition forces an interface $\Gamma$
into $V_{N},$ separating the $\left(  +\right)  $-phase from the $\left(
-\right)  $-phase. Dobrushin has shown that in the thermodynamic limit
$N\rightarrow\infty$ the distribution of $\Gamma$ goes to a proper limit (in
contrast with the 2D case). This limit describes the behavior of the surface
separating the $\left(  +\right)  $- and the $\left(  -\right)  $-phases. His
method of analysis was what is now called the cluster expansion, based on
Pirogov-Sinai Contour Functional theory. Later on, this approach was applied
to other discrete models in \cite{HKZ, CK, GG}.

The question of coexistence of phases for systems with continuous symmetry was
addressed in \cite{FP}. It was found there that the analogous states for the
$XY$-model do not exist, and that the surface tension between two magnetized
phases vanishes. Other systems were not studied in the literature. There are
probably two reasons for that:

\begin{enumerate}
\item most systems with continuous symmetry do not display the above
Ising-type rigid interface separating different phases,

\item the Pirogov-Sinai theory (PS) \textquotedblleft does not
work\textquotedblright\ for continuous symmetry systems, while the (only)
alternative method -- the Reflection Positivity (RP) -- works just for
periodic boundary conditions, and therefore one can not handle boundary
conditions of the type $\sigma^{\pm}$ needed in order to create the interface.
\end{enumerate}

In order to illustrate the first point, let us consider the low-temperature 3D
classical $XY$ model, defined by the Hamiltonian%
\begin{equation}
H\left(  \varphi\right)  =-\sum_{\substack{x,y\in\mathbb{Z}^{3},\\\left\vert
x-y\right\vert =1}}\cos\left(  \varphi_{x}-\varphi_{y}\right)  , \label{01}%
\end{equation}
where the spins $\varphi_{\cdot}$ are taking values on the circle
$\mathbb{S}^{1}=\mathbb{R}^{1}\mathrm{\operatorname{mod}}\left(
2\pi\right) .$ As was established in the seminal paper \cite{FSS},
this model has a continuum of low-temperature magnetized phases,
$\left\langle \cdot \right\rangle _{\alpha},$
$\alpha\in\mathbb{S}^{1}$. One can try to create a state of
coexistence of two phases by using the $\left(  \pm\right)
$-boundary condition\textit{ }$\varphi^{\pm},$ which assigns the
value $0$ to spins in the upper half-space and the value $\pi$ in
the lower half-space. However, as the comparison with the Gaussian
case shows, one expects the thermodynamic limit of that state to be
the mixture state, $\frac{1}{2}\left( \left\langle
\cdot\right\rangle _{\pi/2}+\left\langle \cdot\right\rangle
_{3\pi/2}\right)  ,$ with no interface emerging. The $XYZ$ model is
defined by the same Hamiltonian $\left(  \ref{01}\right)  ,$ but the
variables $\varphi $-s are taking values on the sphere
$\mathbb{S}^{2}\in\mathbb{R}^{3},$ and the difference
$\varphi_{x}-\varphi_{y}$ is just the angle between $\varphi_{x}$
and $\varphi_{y}.$ Let $\left(  \psi,\theta\right)  $ be the
\textquotedblleft Euler angles\textquotedblright\ coordinates on
$\mathbb{S}^{2},$ $\psi \in\mathbb{S}^{1},$ $\theta\in\left[
-\frac{\pi}{2},\frac{\pi}{2}\right]  .$ Again, at low temperatures
there are extremal Gibbs states $\left\langle \cdot\right\rangle
_{\left(  \psi,\theta\right)  },$ $\left( \psi,\theta\right)
\in\mathbb{S}^{2}.$ The $\left(  \pm\right) $-boundary condition
$\varphi^{\pm}$ is now the configuration assigning the value
$\theta=\frac{\pi}{2}$ (north pole) to the upper half-space, and
$\theta=-\frac{\pi}{2}$ (south pole) to the lower half-space. We
expect that the corresponding finite-volume state $\left\langle
\cdot\right\rangle _{\varphi^{\pm}}^{N}$ converges weakly, as
$N\rightarrow\infty,$ to the mixture $\int\left\langle
\cdot\right\rangle _{\left(  \psi,\theta=0\right) }~d\psi.$

Still, we believe that Dobrushin states for some systems with continuous
symmetry do exist. One likely example is the so-called non-linear sigma-model,
considered recently in \cite{ES1, ES2}. Its Hamiltonian is given by%
\begin{equation}
H\left(  \varphi\right)  =-\sum_{\substack{x,y\in\mathbb{Z}^{d},\\\left\vert
x-y\right\vert =1}}\left(  \frac{1+\cos\left(  \varphi_{x}-\varphi_{y}\right)
}{2}\right)  ^{p}, \label{223}%
\end{equation}
with $\varphi_{x}\in\mathbb{S}^{1}.$ For $p$ large enough -- i.e. when the
potential well is narrow enough -- this model exhibits the following behavior:
at high temperatures it has unique Gibbs state (the chaotic state). At low
temperatures in 2D it presumably has the Kosterlitz-Thouless phase with
power-law correlation decay, which can be obtained by the methods of the paper
\cite{FS}. At low temperatures in 3D it should have infinitely many ordered
Gibbs states, indexed by magnetization, as the results of \cite{FSS} suggest.
Moreover -- and that is the main result of \cite{ES1} -- there exists a
critical temperature $T_{c}=T_{c}\left(  p,d\right)  ,$ at which we have the
coexistence of the chaotic state and the ordered state(s). (Of
course, all these states are translation-invariant.) The results of \cite{ES1}
are valid in any dimension $d\geq2.$ We believe that in dimension $d=3$ at the
critical temperature $T_{c}$ the system possesses also
non-translation-invariant states, describing the coexistence of ordered states
and chaotic state, with the rigid interface separating them.

The present paper was started as an attempt to prove the above conjecture.
Unfortunately, we are currently unable to complete this program. (Our partial
results in this direction are briefly described at the end of this
introduction.) However, we are able to study the interfaces in some discrete
approximations of the non-linear sigma-model and other models of this type. By
discrete approximation we mean here the following. Let $H\left(
\varphi\right)  =-\sum_{\substack{x,y\in\mathbb{Z}^{3},\\\left\vert
x-y\right\vert =1}}U\left(  \left\vert \varphi_{x}-\varphi_{y}\right\vert
\right)  $ be the Hamiltonian for the continuous spin model, $\varphi_{x}%
\in\mathbb{S}^{1},$ with free measure $d\varphi.$ Then its discrete
approximation is given by the Hamiltonian
\begin{equation}
H\left(  \sigma\right)  =-\sum_{\substack{x,y\in\mathbb{Z}^{3},\\\left\vert
x-y\right\vert =1}}U\left(  \left\vert \sigma_{x}-\sigma_{y}\right\vert
\right)  ,\label{1010}%
\end{equation}
with $\sigma_{x}\in\mathbb{Z}_{q}\subset\mathbb{S}^{1},$ where the group
$\mathbb{Z}_{q}$ is equipped with counting measure. The integer $q$ is the
parameter of the approximation. (One can call the resulting model as
\textit{the} \textit{clock-model,} corresponding to the interaction $U.$)

If the function $U$ has unique nondegenerate minimum on $\mathbb{S}^{1},$ then
the resulting $\mathbb{Z}_{q}$-model at low temperatures is Potts-like, and
thus has properties quite different from the continuous symmetry system. The
situation becomes much more interesting if the \textit{minimum of }$U$
\textit{is degenerate} and, moreover, the \textit{minimal value is attained
along a (small) segment}, while the discretization parameter $q$ is large.
Then the properties of such a system are quite similar to the one with
continuous symmetry. Unlike the Potts model, the ground states of our
Hamiltonian $\left(  \ref{1010}\right)  $ are infinitely degenerate. We
believe that in the 3D case at low temperatures (as well as at zero
temperature) such a model exhibits spontaneous magnetization, while the
truncated correlations decay as a power law. Hopefully one can establish this
conjectured behavior by a suitable version of the infrared bounds. In the 2D
case we believe that \textquotedblleft Mermin-Wagner\textquotedblright%
\ theorem holds, so that the magnetization is zero, even at zero temperature.
We expect the correlation decay to be a power law. Our 2D conjecture at zero
temperature is close in spirit to the results of R. Kenyon \cite{K} on 2D
tilings, while for positive low temperatures its behavior looks to us to be
similar to that of the intermediate phase of the classical clock-model,
established in \cite{FS}. Another model with similar features was considered
by M. Aizenman, \cite{A}.

The methods of the cited papers \cite{ES1, ES2} can be easily
adapted to prove that in dimension $d\geq2$ the structure of the
phase diagram for the Hamiltonian $\left(  \ref{1010}\right)  $,
with the function $U\ $ having deep and $\varepsilon$-narrow well
(possibly with a flat bottom) and $q$ large enough,
has the same features as for the \textquotedblleft very\textquotedblright%
\ non-linear sigma-model: at high temperatures it has unique Gibbs
state, while at low temperatures it has (one or more) Gibbs states
\textquotedblleft with local order\textquotedblright, which means
that the probability of seeing the discrepancy: $\left\vert
\sigma_{i}-\sigma_{j}\right\vert \geq 3\varepsilon$ at two n.n.
sites is small. Moreover, there exists a temperature $T_{c}\left(
q\right)  $ at which the high-T chaotic state coexists with the
low-T locally-ordered state(s).

The main result of the present paper is the rigidity property of the
chaos/order interface once the dimension $d$ is at least $3.$ Namely, we show
that if the two phases are put into coexistence at the transition temperature
$T_{c}\left(  q\right)  $ by applying suitable boundary conditions in a given
volume, then the interface between them is rigid, and its height function
exhibits the long-range order. Since the proof of this result is quite
involved, we will establish it in the present paper only for the simplest
model of the above type, defined below, $\left(  \ref{20}\right) .$

We will now comment on the method we use to study our problem. Presently there
are two techniques to study phase transitions: the Pirogov-Sinai theory and
the method of Reflection Positivity. It seems unlikely that our model can be
treated by PS-theory, since we have here infinite degeneracy of the ground
states and we expect power-law decay of correlations. On the other hand, the
applications of the RP method rely on the study of the states with periodic
boundary conditions. In the phase coexistence regime such a state is not
ergodic, and its ergodic decomposition allows one to study various pure states
-- but only \textit{some states,}since the non-translation-invariant
states do not contribute to the state with periodic boundary conditions.

Notwithstanding the above discussion, our method of proof will be that of
Reflection Positivity. But in order to study the chaos/order interface, we
will use RP not with periodic boundary conditions, but with mixed ones;
namely, we impose periodic boundary conditions only in two (horizontal)
dimensions, and we leave the third (vertical) dimension \textquotedblleft
free\textquotedblright\ to impose fixed spins boundary conditions in the third
dimension. In other words, we consider the cylindric boundary conditions
topology, $\mathbb{T}_{N}^{2}\times\left[  0,L\right]  ,$ and we impose
ordered boundary conditions on the top of the cylinder and chaotic boundary
conditions on its bottom. Of course, the resulting state will be RP only with
respect to reflections in vertical planes, but that will be sufficient for our
purposes. This restricted Reflection Positivity is the main technical
innovation of this paper.

\textbf{Our main result} will be that the so constructed state at $T=T_{c}$
necessarily possesses an interface, separating the ordered and the disordered
phases, which interface is rigid in the sense of \cite{D72}: it has a
well-defined (random) global height, while the deviations from it happen at
any given location with a small probability.

One usual advantage of RP method and the chess-board estimates is that their
technical implementations are usually quite simple, as compared with the
Pirogov-Sinai theory. In this respect we have to note that the
\textit{restricted }RP is already more involved technically and requires a
detailed study of various spatially extended defects, not present in the usual
applications of RP.

Our technique enables one to study also the continuous symmetry
case, and to obtain similar results in a 3D slab
$\mathbb{Z}^{2}\times\left[  0,L\right] :$ with order-disorder
boundary conditions and for a suitable narrow-well interaction $U$
one has the chaos/order rigid interface at the critical temperature
$T_{c}.$ However, the technical limitations of our approach are such
that  the width $\varepsilon$ of the potential well depends on the
width $L$ of the slab, with $\varepsilon\left(  L\right)
\rightarrow0$ as $L\rightarrow\infty.$ Therefore, in contrast to the
discrete symmetry case, we can not take the full thermodynamic limit
$N\rightarrow \infty,$ $L\rightarrow\infty.$ The details will be
published separately,  see \cite{V1}.

\textbf{Other models.} Finally we remark that our technique, applied to the 3D
Ising or Potts models, allows one to obtain simpler proofs of the rigidity of
their interfaces. Indeed, since in these models the ground states are
non-degenerate, our machinery simplifies a lot, and the resulting proofs are
relatively short. We can also treat various 3D real valued random fields. For
example, we can study the double-well case, defined by the Hamiltonian%
\begin{equation}
H\left(  \psi\right)  =\sum_{s}\left(  \psi_{s}^{2}-1\right)  ^{2}%
+\sum_{s,t\text{ n.n. }}\left(  \psi_{s}-\psi_{t}\right)  ^{2}.\label{221}%
\end{equation}
We can show that at low temperatures this system possesses rigid interface
separating the plus-phase, where $\psi\approx+1,$ from the minus-phase, where
$\psi\approx-1.$ Another case of interest is the model with extra local
minimum of the energy, considered in \cite{DS}, where
\begin{equation}
H\left(  \psi\right)  =\sum_{s}\Phi\left(  \psi_{s}\right)  +\sum_{s,t\text{
n.n. }}\left(  \psi_{s}-\psi_{t}\right)  ^{2}.\label{222}%
\end{equation}
Here the potential $\Phi$ has a (unique) global minimum, which is narrow, and
an additional local one, which has to be relatively wide. Then, as it is shown
in \cite{DS}, such a model undergoes a phase transition at some temperature
$T_{cr}\left(  \Phi\right)  ,$ at which temperature one has a coexistence of
the low-energy phase, corresponding to the global minimum, with high entropy
phase, corresponding to the local minimum. In dimension 3 we can show that at
this temperature this model exhibits rigid interface separating the above two phases.

We want to stress that the above stated results for the models $\left(
\ref{221}\right)  $ and $\left(  \ref{222}\right)  $ are technically simpler
than the corresponding statements for the system $\left(  \ref{1010}\right)  $
and its discrete versions. Indeed, while in the models $\left(  \ref{221}%
\right)  $ and $\left(  \ref{222}\right)  $ one has exponential decay of
correlation due to the positive mass of the potential wells, in $\left(
\ref{1010}\right)  $ and its discrete version we expect power law decay. This
is why in the present paper we concentrate on the last model. The
corresponding results for $\left(  \ref{221}\right)  $ and $\left(
\ref{222}\right)  $ will be published separately, \cite{SV}.

The organization of the paper is the following:

The next section contains the definition of the model we study and the
formulation of the main result. Section 3 contains the main steps of the
proof. We introduce there the gas of defects of the interface, and we use
Reflection Positivity and the chess-board estimates to reduce the study of the
local defects to the study of defect sheets. Some defects do not contribute to
the weight of the interface, so to control these we have to glue them in pairs
by means of the gluing transformation. The Sections 4 and 5 contain the needed
combinatorial-energy properties of various defect sheets. The last Section 6
contains the final steps of the proof of our main result.

\section{The Main Result}

In what follows we will consider the 3D lattice model with spins $\sigma_{i}$
taking values in the additive group $\mathbb{Z}_{q}=\mathbb{Z}/q\mathbb{Z}.$
We will equip $\mathbb{Z}_{q}$ with the counting measure. Let $\sigma=\left\{
\sigma_{i}:\sigma_{i}\in\mathbb{Z}_{q},i\in\mathbb{Z}^{3}\right\}  $ be a
configuration of our model. The Hamiltonian of our system is given by
\begin{equation}
H(\sigma)=-\sum_{i\sim j}\mathbb{I}_{\left\vert \sigma_{i}-\sigma
_{j}\right\vert \leq1}, \label{20}%
\end{equation}
where the summation goes over nearest neighbors. Clearly, the interaction and
the Hamiltonian are $\mathbb{Z}_{q}$-invariant. (In terms of Section 1, the
Hamiltonian $\left(  \ref{20}\right)  $ corresponds to the model $\left(
\ref{1010}\right)  $ with interaction having a well of width $\varepsilon
=\frac{3}{q}.$)

Let us define the notion of order:

\begin{definition}
[Ordered bonds]A bond $i\sim j$ of our lattice $\Lambda_{N,L}$ is called
ordered in $\sigma$ iff $\left\vert \sigma_{i}-\sigma_{j}\right\vert \leq1$.
Otherwise it is called disordered.
\end{definition}

Using a technique similar to \cite{ES1, ES2}, one can show that for $q$ large
enough the above model undergoes a first-order phase transition in
temperature. Namely, the following theorem holds:

\begin{theorem}
There exists a temperature $T_{c}=T_{c}\left(  q\right)  ,$ at which the
Hamiltonian $\left(  \ref{20}\right)  $ has at least two Gibbs states: the
ordered state $\left\langle \cdot\right\rangle _{T_{c}}^{o}$ and the
disordered state $\left\langle \cdot\right\rangle _{T_{c}}^{d}.$ They are
characterized by the properties:
\begin{align}
\left\langle \mathbb{I}_{\left\vert \sigma_{i}-\sigma_{j}\right\vert \leq
1}\right\rangle _{T_{c}}^{d}  &  \leq p\left(  q\right)  ,\ \ \label{21}\\
\left\langle \mathbb{I}_{\left\vert \sigma_{i}-\sigma_{j}\right\vert \leq
1}\right\rangle _{T_{c}}^{o}  &  \geq1-p\left(  q\right)  , \label{22}%
\end{align}
where $i,j$ is any bond of $\mathbb{Z}^{3},$ while $p\left(  q\right)  $ goes
to zero as $q\rightarrow\infty.$ (Incidentally, the critical temperature
$T_{c}\left(  q\right)  $ goes to zero as $q\rightarrow\infty$.)
\end{theorem}

{\small NOTE. We believe that in 3D the state }$\left\langle \cdot
\right\rangle _{T_{c}}^{o}${\small is not pure, and is a mixture of }$q$
{\small states with different values of magnetization. \medskip}

The purpose of our work is the study of the interface between the ordered and
disordered phases of the Hamiltonian $\left(  \ref{20}\right)  $ at the
critical temperature $T_{c}\left(  q\right)  $, put into coexistence by
suitable boundary conditions.  The construction of the corresponding
non-translation-invariant states will be discussed in another publication,
\cite{V}.

To study the interfaces we will consider special boxes and we will impose
special boundary conditions, which will force the interface into the box.
Namely, we will take the boxes $\Lambda_{N,L}\subset\mathbb{Z}^{3}:$
\[
\Lambda_{N,L}=\left\{  (x,y,z);0\leq x,y\leq N;\;0\leq z\leq L+1\right\}  ,
\]
and we will impose the \textit{periodic boundary conditions in }$x$\textit{
and }$y$\textit{ directions}. In other words, we think about the box
$\Lambda_{N,L}$ as a product of the torus $\mathbb{T}_{N}$ and a segment. In
what follows we suppose that $N$ is even. The boundary of $\Lambda_{N,L}$ has
two components, and we denote them by
\[
\mathcal{P}^{o}=\Lambda_{N,L}\cap\left\{  z=L+1\right\}  \text{ and
}\mathcal{P}^{d}=\Lambda_{N,L}\cap\left\{  z=0\right\}  .
\]

We will impose boundary conditions on $\mathcal{P}^{o}$ and $\mathcal{P}^{d}$,
which (hopefully) would bring the order-disorder interface into $\Lambda
_{N,L}.$ So we fix a value $s\in\mathbb{Z}_{q},$ and we impose on
$\mathcal{P}^{o}$ the \textit{ordered} boundary condition $\sigma
_{\mathrm{ord}}=\left\{  \sigma_{i,\,j,\,L+1}\equiv s\right\}  .$ We also fix
four values: $s_{00}=0$, $s_{01}=\left[  q/4\right]  $, $s_{10}=\left[
3q/4\right]  $ and $s_{11}=\left[  q/2\right]  $ in $\mathbb{Z}_{q},$ and we
impose the \textit{strongly} \textit{disordered} boundary condition
$\sigma_{\mathrm{disord}}=\left\{  \sigma_{a+2i,\,b+2j,\,0}=s_{ab}%
,a,b=0,1\right\}  $ on $\mathcal{P}^{d}.$ The resulting boundary condition
will be called the \textit{order-disorder b.c.}

In what follows we will be interested in the Gibbs states in $\Lambda_{N,L},$
corresponding to the Hamiltonian $\left(  \ref{20}\right)  ,$ with these
order-disorder b.c. at inverse temperature $\beta.$ They will be denoted by
$\mu_{N,L}^{\beta,q},$ while by $Z_{N,L}^{\beta,q}$ we denote the
corresponding partition function.

To formulate our results we need some more definitions. Let a configuration
$\sigma$ in $\Lambda_{N,L}$ be fixed.

\begin{definition}
[Pure and frustrated cubes]We will call an elementary cube of our lattice
$\Lambda_{N,L}$ frustrated in $\sigma,$ if it has both ordered and disordered
bonds among its (twelve) bonds. Otherwise it will be called pure. Any pure
cube is either ordered or chaotic, in obvious sense.
\end{definition}

The set of all frustrated cubes of $\sigma$ will be denoted by $\mathcal{F}
(\sigma).$

\begin{definition}
[Contours, 3D-interfaces]A connected component of $\mathcal{F}(\sigma)$ is
called a 3D \emph{interface}, iff it separates $\mathcal{P}^{o}$ and
$\mathcal{P}^{d}.$ Otherwise it is called a \emph{contour}.
\end{definition}

\begin{remark} 
Here two cubes are called connected, if they share at least one bond.
\end{remark}

The union of all the 3D interfaces of $\sigma$ will be denoted by $I(\sigma).$
The complement $\Lambda_{N,L}\setminus I(\sigma)$ has several connected
components; each one of them is occupied by a phase -- ordered or chaotic. The
type of the phase in any of these components is defined by the type of the
elementary cube on its inner boundary; inside the components the phases might
have of course frustrated contours.

We need the following topological fact:

\begin{proposition}
[Existence of a 3D-interface]With the order-disorder b.c., defined above, each
configuration has at least one 3D-interface.
\end{proposition}

This obvious claim in fact requires a proof, as was pointed out by G.
Grimmett, \cite{G}. One is given in \cite{GG}, though it also can be deduced
from known results of homotopy theory, see, e.g. \cite{D}.

Now we will define the boundary surface, which rigidity we will prove below:

\begin{definition}
[2D-interface]Let $\sigma$ be a configuration in $\Lambda_{N,L},$ with
order-disorder b.c. imposed. The complement $\Lambda_{N,L}\setminus I(\sigma)$
has several (at least two -- containing $\mathcal{P}^{d}$ and $\mathcal{P}%
^{o}$) connected components. Let us consider all its disordered components.
(There is at least one such component -- the one containing the boundary
$\mathcal{P}^{d}\subset\Lambda_{N,L}.)$ We denote their union by
$\mathcal{D}(\sigma);$ this is the \emph{disordered phase }region. Denote by
$\partial\mathcal{D}$ all the plaquettes which belong both to elementary cubes
in $\mathcal{D}$ and to elementary cubes in $I(\sigma).$ It can have several
connected components. Let $B(\sigma)$ be the union of these components, each
of which separates $\mathcal{P}^{d}$ and $\mathcal{P}^{o}.$ It will be called
the \emph{2D-interface, }or just\emph{ the interface.}

A collection of plaquettes $B$ will be called \emph{admissible} if
$B=B(\sigma)$ for some configuration $\sigma$.
\end{definition}

Let us denote by $\Pi:B(\sigma)\rightarrow\mathcal{P}$ the orthogonal
projection onto the plane $\mathcal{P}=\left\{  z=0\right\}  $. A point
$\bar{M}$ of the surface $B(\sigma)$ will be called \emph{regular}, if the
preimage of its projection $\Pi^{-1}\left(  \Pi\left(  \bar{M}\right)
\right)  $ consists of exactly one point, which is $\bar{M}$ itself. The
plaquette $p$ of $B,$ containing $\bar{M}$ will be then also called regular,
as well as the point $M=\Pi\left(  \bar{M}\right)  \in\mathcal{P}$ and its
plaquette. A \emph{ceiling} is a maximal connected component of regular
plaquettes. We split the complement of ceilings of $B$ into connected
components, which will be called \emph{walls}. Note that all plaquettes of a
ceiling $\mathcal{C}$ necessarily belong to the same horizontal plane
$\left\{  z=h(\mathcal{C})\right\}  $, so the \emph{height of a ceiling}
$h(\mathcal{C})$ is well defined. The height $h\left(  M\right)  $ of the
regular point $M\in\mathcal{P}$ is defined in the obvious way. If the point
$M\in\mathcal{P}$ is not regular, we put $h\left(  M\right)  =\infty$ by
definition. The regular points $M$ of the plane $\mathcal{P}$ also can be
splitted into connected components. Let $R\left(  \sigma\right)
\subset\mathcal{P}$ be the one with the largest area. (If there are several
such, we choose one of them.) The set $R\left(  \sigma\right)  $ will be
called the \emph{rigidity }set of $\sigma.$ The preimage $\mathcal{\bar{C}%
}\left(  \sigma\right)  =\Pi^{-1}\left(  R\left(  \sigma\right)  \right)  $ is
(contained in one of) the largest ceiling of $B\left(  \sigma\right)  .$

Our main result states that, typically, the rigidity set is \textit{very} big:

\begin{theorem}
\mbox{} \label{T}

\begin{itemize}
\item \textbf{Rigidity. }Let $q>q_{0},$ with $q_{0}$ being large enough. Let
our box $\Lambda_{N,L}$ has even width $N$, while the height $L$ does not
exceed $\exp\left\{  N^{2/3}\right\}  .$ Then for every $\beta$
\[
\mu_{N,L}^{\beta,q}\left\{  \frac{\left\vert R\left(  \sigma\right)
\right\vert }{N^{2}}>1-a\left(  q\right)  \right\}  \rightarrow1
\]
as $N\rightarrow\infty,$ for some $a\left(  q\right)  >0,$ with $a\left(
q\right)  \rightarrow0$ as $q\rightarrow\infty.$ In particular, the
surface $B$ has typically only one connected component.

\item \textbf{Long-range order.} The function $h\left(  M\right)  $ is the
long-range order parameter: if $M^{\prime},M^{\prime\prime}$ are two arbitrary
points in $\mathcal{P},$ then the probability of the event
\[
\mu_{N,L}^{\beta,q}\left\{  h\left(  M^{\prime}\right)  =h\left(
M^{\prime\prime}\right)  \text{ and are finite}\right\}  \rightarrow1
\]
as $q\rightarrow\infty,$ uniformly in $N$ and $M^{\prime},M^{\prime\prime}$
and for every $\beta.$
\end{itemize}
\end{theorem}

Of course, for most values of the temperature this result is not very
surprising. Indeed, if $T>T_{cr},$ say, then the box $\Lambda_{N,L}$ will be
filled with disordered phase, while the surface $B(\sigma)$ is pressed to stay
in the vicinity of the $P^{o}$-component of the boundary. Our result is of
real interest precisely at  criticality, since at $T=T_{cr}$ the surface
$B(\sigma)$ stays away from the boundaries of the box $\Lambda_{N,L}$ due to
the entropic repulsion. We expect that at  criticality the location $h\left(
\mathcal{\bar{C}}\left(  \sigma\right)  \right)  $ of the interface
$B(\sigma)$ is distributed approximately uniformly in the segment $\left[
C\ln N,L-C\ln N\right]  .$ The details will be given in \cite{V}.

We would like to comment that the power of the RP method lies in the property
that one can make statements about the behavior at the critical point by
establishing some features for general temperatures. Indeed, it would be very
difficult for us to work precisely at the critical temperature, since we do
not even know its exact value.

The main step towards the proof of rigidity is the control of the fluctuations
of the interface with respect to the optimal flat shape. We thus make the
following definition:

\begin{definition}
Let $B$ be the interface, and $D\subset B$ be any collection of plaquettes. We
define the \emph{weight} of $D$ to be $w(D)=\left\vert D\right\vert
-\left\vert \Pi(D)\right\vert $, where $\left\vert \cdot\right\vert $ is the
number of plaquettes in the collection.
\end{definition}

We have the following estimate:

\begin{theorem}
[Peierls estimate]\label{thm:continuous_peierls} Suppose that $N$ is even.
Then, for all $\beta,L$ and all collections of plaquettes $D$,
\[
\mu_{N,L}^{\beta,q}\left(  B:D\subset B\right)  \leq a^{w(D)},
\]
where $a=a(q)$ goes to $0$ when $q\rightarrow\infty$.
\end{theorem}

\section{Proof of the Theorem \ref{thm:continuous_peierls}}

\label{sec:toy}

\subsection{Settings for reflection positivity, construction of the blobs}

In order to set the framework for reflection positivity, we consider the
system as a spin-system on the 2-dimensional torus $\mathbb{T}_{N}$, where at
each site of $\mathbb{T}_{N}$ we have a random variable taking values in
$\left(  \mathbb{Z}_{q}\right)  ^{L}$ (we recall that $\mathbb{Z}%
_{q}=\mathbb{Z}/q\mathbb{Z}$).

It is straightforward to see that $\mu_{N,L}^{\beta,q}$ is reflection positive
with respect to the group generated by the reflections in the lines passing
through the sites of the torus, see any of the RP papers \cite{FL, FILS}, or
the review \cite{S}.

Let $p\subset\mathbb{T}_{N}$ be any plaquette. Its full preimage $c=\Pi
^{-1}\left(  p\right)  \subset\Lambda_{N,L}$ will be called a column. The set
of all columns will be denoted by $C_{N}.$ Any \emph{horizontal} plaquette
$\Lambda_{N,L}$ belongs to a well defined column, but for (some)
\emph{vertical} plaquettes we will make a $\sigma$-dependent choice. We assume
the following convention: let $P$ be a vertical plaquette, separating a
frustrated cube of configuration $\sigma$ from a pure disordered one; then we
say that $P$ belongs to the column containing the frustrated cube but not to
the column containing the pure one. Now for any column $c\in C_{N}$, we define
$B_{c}=$ $B_{c}\left(  \sigma\right)  $ to be the set of plaquettes of
$B\left(  \sigma\right)  $ contained in $c$.

\begin{definition}
We define the blobs of $\sigma$ in $c$ to be the connected components of
$B_{c}$. We will denote by $\mathfrak{B}(B_{c})=(b_{1},\ldots,b_{r})$ the set
of blobs in the column $c$ for the collection $B\left(  \sigma\right)  $,
enumerated upwards.
\end{definition}

\subsection{Application of the chessboard estimate}

In the first three subsections of this section we will reduce the Peierls
estimate -- the estimate of a local event, see $\left(  \ref{81}\right)  ,$ --
to an estimate of a global event $\left(  \hat{\pi}_{\tau}\right)  _{c}^{N}$,
see $\left(  \ref{014}\right)  .$ The remaining two subsections describe the
splitting of $\left(  \hat{\pi}_{\tau}\right)  _{c}^{N}$ into defects and
their pairing.

\medskip

Let $\sigma$ be some configuration. We distinguish several kinds of blobs in
$\mathfrak{B}(B_{c})=(b_{1},\ldots,b_{r}),$ as we move upwards. The blob
$b_{i}$ has:

\begin{itemize}
\item type $h-$ ($h+$), if $b_{i}$ begins (ends), as one ascends, with a
horizontal plaquette, the rest being vertical; if $b_{i}$ consists of just a
single plaquette, then it is of type $h-$ ($h+$) if the cube below (above) it
is pure disordered;

\item type $h-+$, if $b_{i}$ begins \emph{and}
ends with a horizontal plaquette, the rest being vertical (in that case the
first cubes above it and below it have to be pure disordered);

\item type $v$: $b_{i}$ is a pack of vertical facets.
\end{itemize}

Note that because of the convention we took for vertical plaquettes, there are
no other cases. Moreover, from bottom to top we have the following rules:

\begin{itemize}
\item there exists at least one signed blob, and the blob-signs are alternating;

\item the first and the last signs are $-,$

\item the first signed blob after a $v$-blob is of the type $h+.$
\end{itemize}

\begin{remark}
If $B_{c}$ is made of exactly one horizontal plaquette, there is only one blob
in $c$, and it is of type $h-$. This blob is called trivial.
\end{remark}

\subsubsection{Defining Defects}

Let us now consider the set $F\left(  \sigma\right)  $ of all frustrated
cubes, attached to $B\left(  \sigma\right)  .$ We will denote by $F_{c}\left(
\sigma\right)  $ the intersection $F\left(  \sigma\right)  \cap c.$ Let
$C_{N}\left(  \sigma\right)  $ be the set of all columns $c,$ such that
$B_{c}$ contains at least two plaquettes. For $c\in C_{N}\left(
\sigma\right)  $ let $F_{i}$ $\subset F_{c}\left(  \sigma\right)  ,$
$i=1,...,r^{\prime}$ be connected components of $F_{c}\left(  \sigma\right)
$. These segments of frustrated cubes will be called \emph{defects} of
$\sigma.$ Now, every blob $b_{j}$ is contained in some defect $F_{i},$ but
since some $F_{i}$-s can contain several blobs, we have $r^{\prime}\leq r.$
The set of all defects of $\sigma$ is denoted by $\pi\left(  \sigma\right)  ,$
while $\pi_{c}\left(  \sigma\right)  \subset\pi\left(  \sigma\right)  ,$ $c\in
C_{N}\left(  \sigma\right)  $ will be those belonging to the column $c.$

Our immediate goal will be the proof of the following

\begin{proposition}
\label{P} Let $D\subset\Lambda_{N,L}$ be any collection of cubes. Then
\begin{equation}
\mu_{N,L}^{\beta,q}\left(  \sigma:D\subset\pi\left(  \sigma\right)  \right)
\leq a^{\left\vert D\right\vert -\left\vert \Pi\left(  D\right)  \right\vert
}, \label{81}%
\end{equation}
where $a=a(q)$ goes to $0$ when $q\rightarrow\infty$.
\end{proposition}

The Peierls estimate evidently follows from this.

The rough idea of proving the Proposition \ref{P} is the following.
We will try to show that the cost of having a defect with $k$
frustrated cubes is of the order of $c^{k},$ $c<1.$ This is indeed
true, and we will show that for all defects with $k\geq2$ the price
behaves as $c^{k-1}.$ However, for some defects with $k=1$ there is
no price to pay at all, due to our choice of boundary conditions,
which force the interface - and hence the defects - into the system.
We will show then that if there are several such problematic defects
-- i.e. defects with $k=1$ -- then one can pair them, and extract
the cost contribution of the order of $c$ for every pair. This will
be enough for our purposes.

{\small NOTE. The reader who would like to understand first the easy part of
the proof -- the one dealing with non-problematic defects -- can go after the
Definition \ref{probl} below straight to the Section \ref{NP}. }

To implement the above strategy we need to impose some more structure on the
defects. First of all, we define their signs. Namely, each defect $F$
contains several blobs. Let us add all the signs of all the blobs in $F.$ The
resulting sign will be called the sign of $F,$ \textrm{sgn}$\left(  F\right)
.$ It takes values $+,-$ or $0.$ Since in the string of blobs in $c$ the signs
are alternating, the sign of $F$ is well defined.

We will also need some information about the vicinity of the defects. So we
will spatially extend the defects, fixing to a certain extent the
configuration at their ends. Then, of course, we will have to perform the
summation over all extensions. In the process of extension some defects might
coagulate into a single bigger defect, in which case we always will treat the
result as a single defect.

\subsubsection{Extending Defects}

Here we will describe the process of extending the defects. The extension will
depend on $\sigma,$ of course.

On the first step we extend each defect $F_{j}\subset c$ to a longer segment
of cubes $\phi_{1}\left(  F_{j}\right)  ,$ $F_{j}\subset\phi_{1}\left(
F_{j}\right)  \subset c,$ which is a minimal segment containing $F_{j},$ which
contains, apart from $F_{j},$ only frustrated cubes, except two end-cubes,
which are pure cubes. (The added cubes need not touch the interface.) In the
case that the defect $F$ is attached to the boundary of $\Lambda_{N,L}$, the
extended defect has at most one pure end-cube. Evidently, the operation
$\phi_{1}$ is well-defined. It can happen that some resulting segments
$\phi_{1}\left(  F_{j}\right)  $ and $\phi_{1}\left(  F_{j^{\prime}}\right)  $
have an elementary pure cube in common. In that case we merge them into a
single defect: we will consider the connected components of the family
$\left\{  \phi_{1}\left(  F_{i}\right)  \right\}  $, and by a slight abuse of
terminology we still call the resulting segments defects (or extended
defects). The sign of the merger is defined to be the sum of the constituents.
Now any two defects have no cubes in common (though they can share a facet).
From now on we will deal exclusively with extended defects, so in what follows
we will omit the symbol $\phi_{1}$ and will write just $F$ for the extended defects.

We also \emph{fix the nature of every bond in the defect}, i.e. whether the
bond is ordered or disordered.

\begin{figure}
\centering
\includegraphics[width=0.9\textwidth]{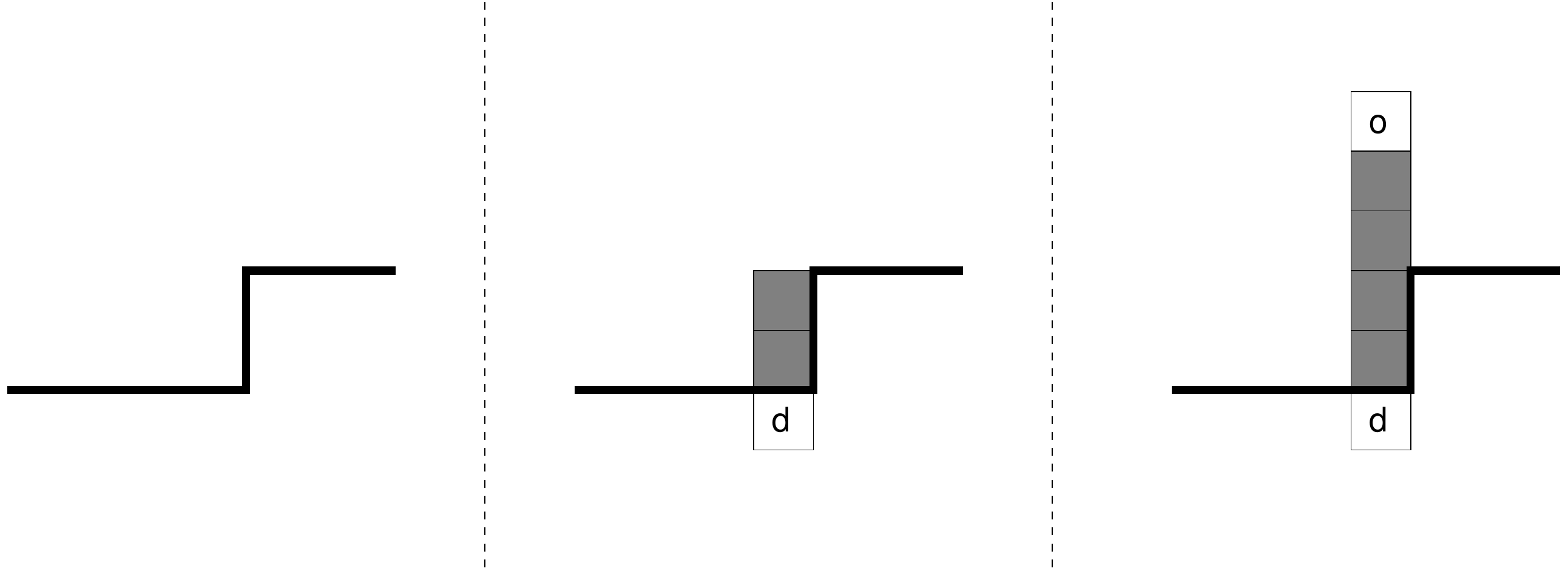}
\caption{A piece of interface, which generates a (non-problematic)
defect; frustrated cubes are shaded. The third picture shows one
possible outcome of the extension of the defect.}
\label{fig=non_problematic}
\end{figure}

\begin{figure}
\centering
\includegraphics[width=0.9\textwidth]{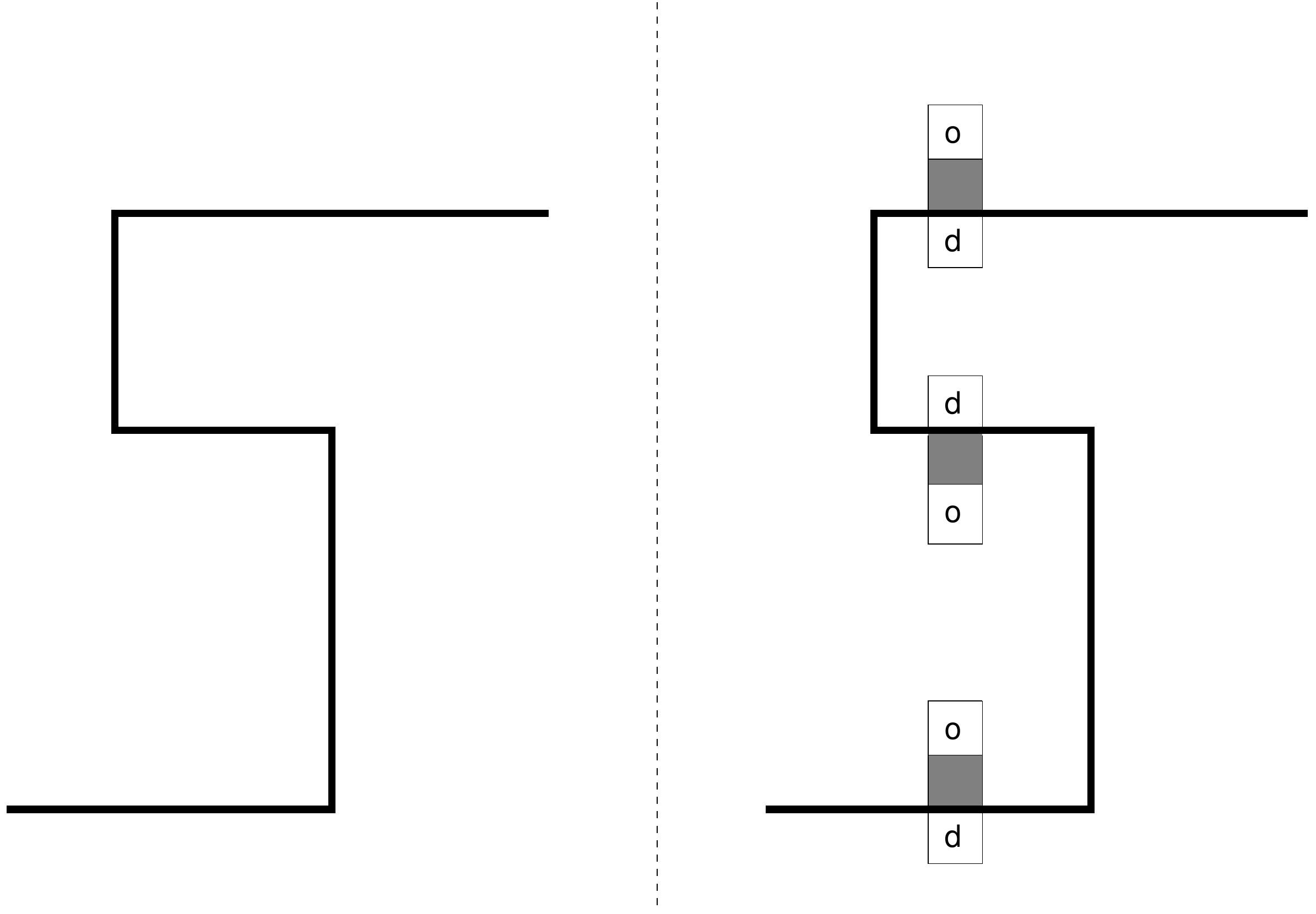}
\caption{A piece of interface, which generates three problematic
defects.} \label{fig=problematic}
\end{figure}

\begin{definition}
[Problematic defects]\label{probl} Among the defects we single out those with
the property that every bond not belonging to the two end-cubes is disordered.
(Note that at least one of these end-cubes has then to be ordered.) If this
defect is signed, it was built from a blob consisting of just one horizontal
plaquette; if it is not signed, it was built from the coagulation of two
consecutive signed blobs, both consisting of just one horizontal plaquette. In
both cases these defects will be called \emph{problematic. } If both end-cubes
of a problematic defect are ordered, the defect consists of 5 cubes, 3 of
which are pure; if only one end-cube is ordered, the defect consists of 3
cubes, 2 of which are pure.

Other defects, which will be called exceptional problematic defects, appear
among defects attached to the bottom (disordered) boundary. Such a defect is
called e-problematic, if it has the following three properties:

1. It consists from one or two frustrated cubes, followed by one ordered cube
at the top of the defect,

2. The bottom cube has at least 3 vertical disordered bonds,

3. The corresponding blob consists of exactly one plaquette, which is the
horizontal plaquette at the bottom of the box $\Lambda_{N,L}.$

In particular, any e-problematic defect has sign $\left(  -\right)  .$

All other defects will be called \emph{non-problematic}.
\end{definition}
See Figures \ref{fig=non_problematic},\ref{fig=problematic} for
(two-dimensional!) sketches of non-problematic and problematic
defects.

Thus we have assigned to every configuration $\hat{\sigma}$ with $B\left(
\hat{\sigma}\right)  =B\left(  \sigma\right)  =B$ and to every column $c\in
C_{N}\left(  \sigma\right)  $ the extension $\hat{\pi}_{c}$ of the initial set
$\pi_{c}\left(  \sigma\right)  ,$ including into the extension the
order-disorder specification of every bond of $\hat{\pi}_{c}.$ The set of all
possible extensions $\hat{\pi}$ of $\pi$ will be denoted by $\mathcal{E}%
\left(  \pi\right)  .$ Evidently, we have the partition
\[
\left\{  \sigma:B\left(  \sigma\right)  =B,\pi\left(  \sigma\right)
=\pi\right\}  =\cup_{\hat{\pi}\in\mathcal{E}\left(  \pi\right)  }\left\{
\sigma:B\left(  \sigma\right)  =B,\hat{\pi}\left(  \sigma\right)  =\hat{\pi
}\right\}  ,
\]
so%

\[
\mu_{N,L}^{\beta,q}\left(  \sigma:\pi\left(  \sigma\right)  =\pi\right)
\mathbf{=}\sum_{\hat{\pi}\in\mathcal{E}\left(  \pi\right)  }\mu_{N,L}%
^{\beta,q}\left(  \sigma:\hat{\pi}\left(  \sigma\right)  =\hat{\pi}\right)  .
\]
We will also use the notation $\sigma\in\hat{\pi},$ in the obvious sense. A
straightforward combinatorial counting of the possible extensions of a given
defect shows that to prove $\left(  \ref{81}\right)  $ it is enough to show
that%
\begin{equation}
\mu_{N,L}^{\beta,q}\left(  \sigma:\hat{\pi}\left(  \sigma\right)  =\hat{\pi
}\right)  \leq a^{\left\Vert \hat{\pi}\right\Vert -\left\vert \Pi\left(
\hat{\pi}\right)  \right\vert }\label{0110}%
\end{equation}
(with some smaller $a$), where $\left\Vert \hat{\pi}\right\Vert $ is the
number of frustrated cubes in $\hat{\pi},$ and $\left\vert \Pi\left(
\hat{\pi}\right)  \right\vert $ is the number of plaquettes in the projection
$\Pi\left(  \hat{\pi}\right)  .$

\subsubsection{Fixing the boundary conditions for defects \label{types}}

The last phase of fixing the environment of the defect consists in fixing the
type of the configuration on ordered plaquettes $P$ at the boundaries of the
defect. If the plaquette $P=(x,y,z,w)$ is fully ordered, with $\sigma
(x)-\sigma(y)=a$, $\sigma(y)-\sigma(z)=b$, $\sigma(z)-\sigma(w)=c$, and
$\sigma(w)-\sigma(x)=d$, we say that $\sigma$ is of $(a,b,c,d)$-type on $P$;
we notice that since $a,b,c,d\in\left\{  -1,0,1\right\}  $, there are at most
$3^{4}=81$ possible ordered types for $\sigma$ on $P$. We denote by
$\mathcal{T}$ the set of all possible types.

Each defect $F$ is delimited by two horizontal plaquettes: the top one,
$F^{t},$ and the bottom one, $F^{b}$ ; we define $\partial F=F^{t}\cup F^{b}$.
Each of these plaquettes can be either fully ordered or fully disordered; we
denote by $\partial^{o}F\subset\partial F$ the ordered plaquettes of $\partial
F$ (the subset $\partial^{o}F$ depends on $\hat{\pi}$).

For every collection $\hat{\pi}$ of extended defects, $\hat{\pi}\in
\mathcal{E}\left(  \pi\right)  ,$ we define $\partial\hat{\pi}=\cup_{F\in
\hat{\pi}}\partial F$ and $\partial^{o}\hat{\pi}=\cup_{F\in\hat{\pi}}%
\partial^{o}F$. We refine the partition $\mathcal{E}\left(  \pi\right)  $ by
specifying the types the configuration $\sigma$ has on every plaquette from
the set $\partial^{o}\hat{\pi}:$ if $\tau\in\mathcal{T}^{\partial^{o}\hat{\pi
}}$, we define
\[
\hat{\pi}_{\tau}=\left\{  \sigma\in\hat{\pi}:\forall P\subset\partial^{o}%
\hat{\pi},\,\sigma\text{ is of type }\tau(P)\text{ on }P\right\}  ,
\]
so
\[
\hat{\pi}=\bigcup_{\tau\in\mathcal{T}^{\partial^{o}\hat{\pi}}}\hat{\pi}_{\tau
}.
\]

We notice that for any column $c\in C_{N}\left(  \sigma\right)  $, containing
non-trivial blob, we have for the corresponding defect, that the number of
plaquettes $\left\vert \partial^{o}\hat{\pi}_{c}\right\vert \leq3\left(
\left\Vert \hat{\pi}_{c}\right\Vert -1\right)  $ (with equality iff $\hat{\pi
}_{c}$ consists of two problematic defects -- the first one with
order-disorder b.c., the second with order-order b.c. and with one frustrated
cube each). In particular,
\begin{equation}
\mu_{N,L}^{\beta,q}\left(  \sigma:\hat{\pi}\left(  \sigma\right)  =\hat{\pi
}\right)  =\sum_{\tau\in\mathcal{T}^{\partial^{o}\hat{\pi}}}\mu_{N,L}%
^{\beta,q}\left(  \hat{\pi}_{\tau}\right)  \leq\left(  81\right)  ^{3\left(
\left\Vert \hat{\pi}\right\Vert -\left\vert \Pi\left(  \hat{\pi}\right)
\right\vert \right)  }\sup_{\tau\in\mathcal{T}^{\partial^{o}\hat{\pi}}}%
\mu_{N,L}^{\beta,q}\left(  \hat{\pi}_{\tau}\right)  .\label{012}%
\end{equation}
(The estimate $\left(  \ref{012}\right)  $ is helpful in the discrete case,
since the reflected event $\left(  \hat{\pi}_{\tau}\right)  _{c}^{N}$ (see
below) has a relatively simple structure. This is not so in the continuous
symmetry case.)

In the following $\hat{\pi}\in\mathcal{E}\left(  \pi\right)  $ and $\tau
\in\mathcal{T}^{\partial^{o}\hat{\pi}}$ will be fixed, and we will estimate
from above the $\mu_{N,L}^{\beta,q}$-probability of the event $\left\{
\sigma\in\hat{\pi}_{\tau}\right\}  $. We have%

\begin{equation}
\left\{  \sigma\in\hat{\pi}_{\tau}\right\}  =\cap_{c\in C_{N}\left(
\sigma\right)  }\left\{  \sigma\in\left(  \hat{\pi}_{\tau}\right)
_{c}\right\}  , \label{31}%
\end{equation}
where the event $\left(  \hat{\pi}_{\tau}\right)  _{c}$ consists of
configurations $\sigma$ which in the column $c$ have their pattern of extended
defects equal to $\hat{\pi}_{c},$ while their restriction to the plaquettes
$\partial^{o}\hat{\pi}\cap c$ have types defined by $\tau_{c}\equiv
\tau\Bigm|_{\partial^{o}\hat{\pi}\cap c}.$

The application of the chess-board estimate (see \cite{FILS}, relation (4.4))
reduces the problem of getting the upper bound for the probability $\mu
_{N,L}^{\beta,q}\left\{  \sigma\in\hat{\pi}_{\tau}\right\}  $ to that for all
probabilities $\mu_{N,L}^{\beta,q}\left\{  \sigma\in\left(  \hat{\pi}_{\tau
}\right)  _{c}^{N}\right\}  ,$ $c\in C_{N}\left(  \sigma\right)  ,$ where the
event $\left(  \hat{\pi}_{\tau}\right)  _{c}^{N}$ is the result of applying
multiple reflections to $\left(  \hat{\pi}_{\tau}\right)  _{c}.$ (The
reflected event $\left(  \hat{\pi}_{\tau}\right)  _{c}^{N}$ is described in
details in the following subsection.) Namely, the chess-board estimate claims
that%
\begin{equation}
\mu_{N,L}^{\beta,q}\left\{  \sigma\in\hat{\pi}_{\tau}\right\}  \leq\prod
_{c}\left[  \mu_{N,L}^{\beta,q}\left\{  \sigma\in\left(  \hat{\pi}_{\tau
}\right)  _{c}^{N}\right\}  \right]  ^{\frac{1}{N^{2}}}.\label{013}%
\end{equation}
We will prove that uniformly in $\tau$
\begin{equation}
\mu_{N,L}^{\beta,q}\left(  \left(  \hat{\pi}_{\tau}\right)  _{c}^{N}\right)
\leq a^{N^{2}\left(  \left\Vert \hat{\pi}_{c}\right\Vert -1\right)
},\label{014}%
\end{equation}
provided that \textquotedblleft the interface $B$ is not regular in the column
$c$ \textquotedblright; that means that for any $\sigma\in\left(  \hat{\pi
}_{\tau}\right)  _{c}$ the collection of blobs $\mathfrak{B}(B_{c}%
)=(b_{1},\ldots,b_{r})$ of the interface $B\left(  \sigma\right)  $ in the
column $c$ for the collection $B\left(  \sigma\right)  $ is not just one
trivial blob. (We do not care for the situation with the trivial blob, since
it does not contribute to $\left(  \ref{81}\right)  $ anyway.) We will call
such patterns non-trivial. Then $\left(  \ref{014}\right)  ,$ $\left(
\ref{013}\right)  $ and $\left(  \ref{012}\right)  $ imply the relation
$\left(  \ref{0110}\right)  .$

\subsubsection{Description of the reflected event $\left(  \hat{\pi}_{\tau
}\right)  _{c}^{N}$}

The column $c$ is now fixed. The event $\left(  \hat{\pi}_{\tau}\right)  _{c}$
consists of collection of (extended) defects $F_{1},F_{2},\ldots,F_{s}$ in the
column $c,$ each of these equipped with a boundary condition $\tau_{i}%
\in\mathcal{T}^{\partial^{o}F_{i}}.$ Let the slab $\Lambda_{i}=\left\{
(x,y,z);\;0\leq x,y\leq N,\,a_{i}\leq z\leq b_{i}\right\}  $ be the smallest
one containing the defect $F_{i}$. The event $\sigma\in\left(  \hat{\pi}%
_{\tau}\right)  _{c}^{N}$ happens if the following two conditions hold:

\begin{itemize}
\item in every column $c^{\prime}$ the pattern of order/disorder bonds of
configuration $\sigma$ agrees with $\theta_{c,c^{\prime}}\left(  F_{1}%
,F_{2},\ldots,F_{s}\right)  ,$ where $\theta_{c,c^{\prime}}$ is any
composition of the reflections in the lines passing through the sites of the
torus, which takes $c$ to $c^{\prime},$

\item on every ordered plane $z=a_{i}$ (resp. $z=b_{i}$)$,$ $i=1,...,s,$ the
configuration $\sigma$ is of \textquotedblleft reflected\textquotedblright%
\ type $\left(  \tau_{i}^{b}\right)  ^{N}$ (resp. $\left(  \tau_{i}%
^{t}\right)  ^{N}$)$,$ where in column $c^{\prime}$ the type $\left(  \tau
_{i}^{b}\right)  ^{N}$ is defined to be $\theta_{c,c^{\prime}}\left(  \tau
_{i}^{b}\right)  $ (resp. $\theta_{c,c^{\prime}}\left(  \tau_{i}^{t}\right)  $).
\end{itemize}

We denote by $F_{i}^{N}$ the repeated reflection of the defect $F_{i},$ i.e.
$F_{i}^{N}=\cap_{c^{\prime}}\theta_{c,c^{\prime}}\left(  F_{i}\right)  .$ It
is a pattern of order/disorder bonds in $\Lambda_{i}.$ We put $L_{i}%
=b_{i}-a_{i}-1$, and we define $m_{i}$ to be the number of frustrated cubes in
$F_{i}.$ Since every point $\left(  x,y,z\right)  $ with $a_{i}<z<b_{i}$
belongs to at least one frustrated cube of $F_{i}^{N},$ we have
\begin{equation}
L_{i}\leq2m_{i}, \label{72}%
\end{equation}
which will be of importance later. The complement $\Lambda_{N,L}%
\setminus\left(  \cup_{i=1}^{s}\Lambda_{i}\right)  $ is a collection of slabs
$\Lambda_{i}^{\prime}=\left\{  (x,y,z);\;0\leq x,y\leq N,\,b_{i}\leq z\leq
a_{i+1}\right\}  $, $i=0,...,s,$ with the conventions that $b_{0}=0$ and
$a_{s+1}=L+1$.

We now fix the values $\eta$ of the configuration $\sigma$ on $\partial\left(
\hat{\pi}_{\tau}\right)  _{c}^{N}$, i.e. on each plane $z=a_{i}$ or $z=b_{i}$.
The set of $\eta$-s which are compatible with $\left(  \hat{\pi}_{\tau
}\right)  _{c}^{N}$ is denoted by $\mathcal{B}\left(  \left(  \hat{\pi}_{\tau
}\right)  _{c}^{N}\right)  $. We choose some $\eta\in\mathcal{B}\left(
\left(  \hat{\pi}_{\tau}\right)  _{c}^{N}\right)  $, and define
\[
\left(  \hat{\pi}_{\tau}\right)  _{c}^{N}(\eta)=\left(  \hat{\pi}_{\tau
}\right)  _{c}^{N}\cap\left\{  \sigma\Bigm|_{\partial\pi_{c}^{N}}%
=\eta\right\}  .
\]
We obviously have%
\begin{equation}
\mu_{N,L}^{\beta,q}\left(  \left(  \hat{\pi}_{\tau}\right)  _{c}^{N}\right)
=\sum_{\eta\in\mathcal{B}\left(  \left(  \hat{\pi}_{\tau}\right)  _{c}%
^{N}\right)  }\mu_{N,L}^{\beta,q}\left(  \left(  \hat{\pi}_{\tau}\right)
_{c}^{N}(\eta)\right)  . \label{015}%
\end{equation}
Uniformly in $\tau,\eta$, we will get an estimate on $\mu_{N,L}^{\beta
,q}\left(  \left(  \hat{\pi}_{\tau}\right)  _{c}^{N}(\eta)\right)  $.

We will denote by $\eta_{i}^{b}$ (resp. $\eta_{i}^{t}$) the restriction of
$\eta$ to the plane $z=a_{i}$ (resp. $z=b_{i}$). Clearly, the partition
function $Z_{N,L}^{\beta,q}\left(  \left(  \hat{\pi}_{\tau}\right)  _{c}%
^{N}(\eta)\right)  ,$ computed over the set $\left\{  \sigma\in\left(
\hat{\pi}_{\tau}\right)  _{c}^{N}(\eta)\right\}  ,$ factors:
\begin{equation}
Z_{N,L}^{\beta,q}\left(  \left(  \hat{\pi}_{\tau}\right)  _{c}^{N}%
(\eta)\right)  =\prod_{i=1}^{s}Z_{\Lambda_{i}}^{\eta_{i}^{b},\eta_{i}^{t}%
}\left(  F_{i}^{N}\right)  \prod_{i=0}^{s}Z_{\Lambda_{i}^{\prime}}^{\eta
_{i}^{t},\eta_{i+1}^{b}}, \label{51}%
\end{equation}
where the superscripts in the partition functions denote the corresponding
boundary conditions for slabs (with the convention that $\eta_{0}^{t}%
=\sigma_{\mathrm{disord}},$ $\eta_{s+1}^{b}=\sigma_{\mathrm{ord}}$), while the
presence of arguments $F_{i}^{N}$ describe the corresponding periodic
order-disorder pattern of bonds. (We note for clarity that it can happen that
$b_{i}=a_{i+1}$ for some $i,$ in which case the slab $\Lambda_{i}^{\prime}$
degenerates to a plane, and the partition function $Z_{\Lambda_{i}^{\prime}%
}^{\eta_{i}^{t},\eta_{i+1}^{b}}$ is taken over the empty set; we put it to be
$1$ by definition.)

Our goal is now to prove that
\begin{equation}
\prod_{i=1}^{s}Z_{\Lambda_{i}}^{\eta_{i}}\left(  F_{i}^{N}\right)  \leq
a^{N^{2}\left[  \left(  \sum_{i=1}^{s}m_{i}\right)  -1\right]  }\prod
_{i=1}^{s}Z_{\Lambda_{i}}^{\eta_{i}}, \label{016}%
\end{equation}
where $m_{i}$ is the number of frustrated cubes in $F_{i}$, and we use the
shorthand notation $Z_{\Lambda_{i}}^{\eta_{i}}\left(  F_{i}^{N}\right)  \equiv
Z_{\Lambda_{i}}^{\eta_{i}^{b},\eta_{i}^{t}}\left(  F_{i}^{N}\right)  ,$
$Z_{\Lambda_{i}}^{\eta_{i}}\equiv Z_{\Lambda_{i}}^{\eta_{i}^{b},\eta_{i}^{t}%
}.$ Since, obviously,
\[
\sum_{\eta\in\mathcal{B}\left(  \left(  \hat{\pi}_{\tau}\right)  _{c}%
^{N}\right)  }\frac{\prod_{i=1}^{s}Z_{\Lambda_{i}}^{\eta_{i}}\prod_{i=0}%
^{s}Z_{\Lambda_{i}^{\prime}}^{\eta_{i}^{t},\eta_{i+1}^{b}}}{Z_{N,L}^{\beta,q}%
}\leq1,
\]
the relations $\left(  \ref{016}\right)  $ and $\left(  \ref{015}\right)  $
imply $\left(  \ref{014}\right)  .$

We can easily deal with each non-problematic defect $F_{i}$, and we will show
that they satisfy the estimate:
\begin{equation}
Z_{\Lambda_{i}}^{\eta_{i}}\left(  F_{i}^{N}\right)  \leq a^{m_{i}N^{2}}\cdot
Z_{\Lambda_{i}}^{\eta_{i}}. \label{017}%
\end{equation}
However, no reasonable estimate can be obtained for a single problematic
defect. To produce the cost factor needed, we will have to treat the
problematic defects in pairs, and we will produce a factor $a^{2N^{2}}$ for
every such pair.

\begin{figure}
\centering
\includegraphics[width=.7\textwidth]{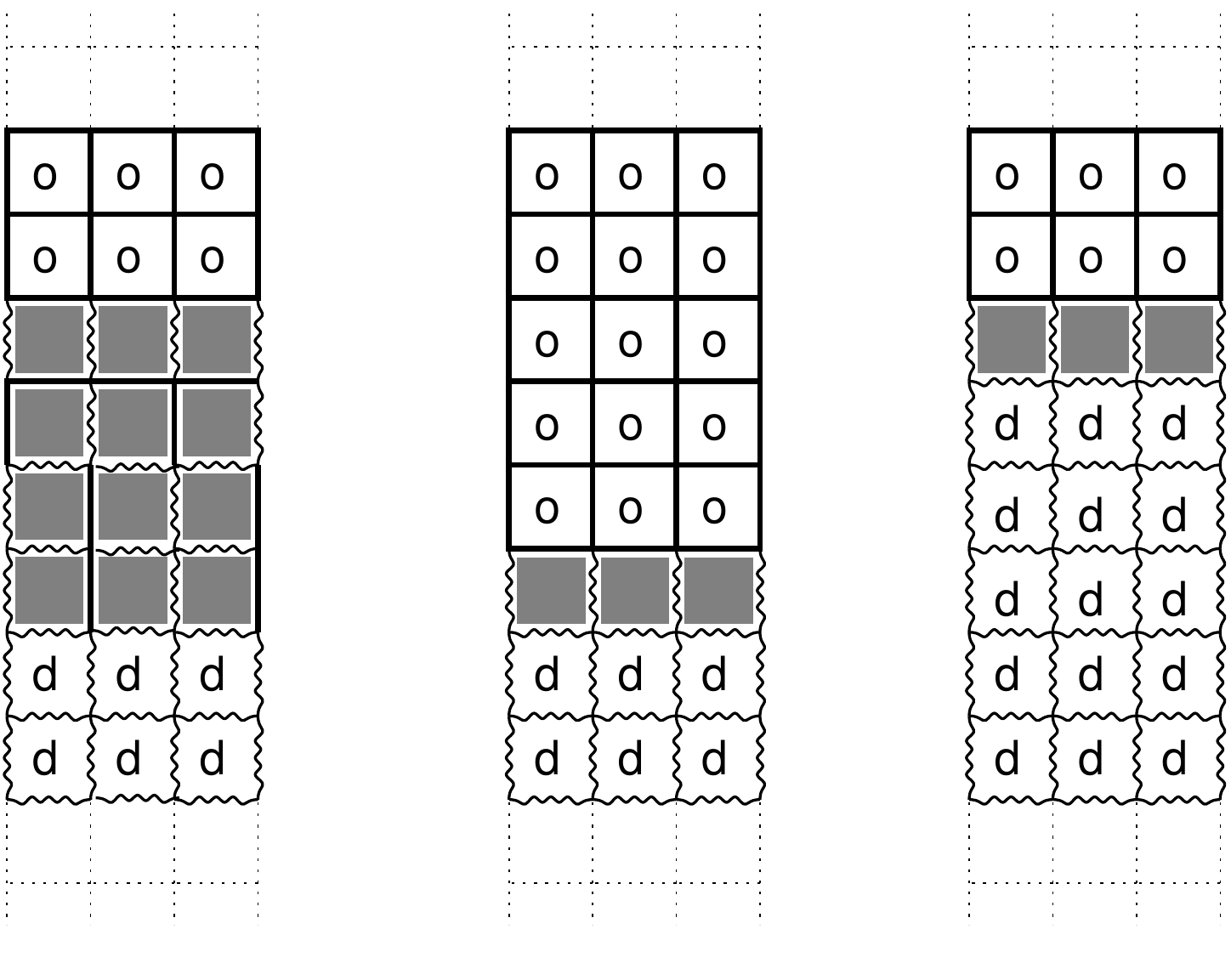}
\caption{From left to right: a defect sheet of width four and its replacements at low and high temperatures. Ordered (disordered) bonds are represented by straight (wriggled) segments.}
\label{fig=heuristics}
\end{figure}

Let us explain the heuristics behind the above claim. Consider for example a
non-problematic defect, which, in ascending order, has the following pattern
of cubes (see Figure \ref{fig=heuristics}):%
\[
\left(  ...,d,d,f,f,f,f,o,o,...\right)  ,
\]
which means that we consider a defect sheet of width 4, sandwiched between the
disordered and ordered phases. We will show in the Section \ref{NP} that the
replacement of it by one of the two following narrower defect sheets:%
\[
\left(  ...,d,d,d,d,d,f,o,o,...\right)
\]
or%
\[
\left(  ...,d,d,f,o,o,o,o,o,...\right)
\]
leads to the increase of the probability.
Which of the last two patterns gives the increase needed depends on the temperature;
in the high temperature region the first scenario (the advance of disorder) wins over the frustration strip,
while at low temperatures the second one takes over the frustration. Moreover
-- and that is of crucial importance -- the two temperature regions are
intersecting, and at the common temperature each of the two scenarios gets a
higher probability than the thick frustration sheet. Note also, that the
frustration sheet can not disappear completely: in every column there should
be at least one frustrated cube between the ordered and the disordered phase,
which is the reason for the problematic defects to be treated separately.

In more details, our strategy will be the following: we consider all signed
defects in the column $c$, which from now on will have their special notation:
$G_{1},G_{2},...,G_{2k-1}.$ Note that we always have an odd number of them;
moreover, their signs alternate, with \textrm{sgn}$\left(  G_{1}\right)
=\left(  -\right)  $. Some of $G_{i}$-s can be problematic. The remaining
neutral defects will be denoted by $H_{1},H_{2},\ldots,H_{l}$; some of them
can also be problematic. We pair signed defects as follows: $\left(
G_{1},G_{2k-1}\right),\left(  G_{2},G_{2k-2}\right),\ldots$ while neutral
defects are paired in the following way: $\left(  H_{1},H_{2}\right), \left(  H_{3},H_{4}\right)
,\ldots$ If $l$ is odd, we finally pair the remaining neutral defect $H_{l}$
with $G_{k}$; if $l$ is even, the defect $G_{k}$ is left unpaired. Notice that
the two paired signed defects have the same sign. Note also that for a
non-trivial pattern it can not happen that we have just one defect of
problematic or e-problematic type.

The above pairing will be essential for us only when both defects in the pair
are problematic -- i.e. when we have a \emph{problematic pair}. In that case
we will treat them together via gluing construction, explained below. The
pairing of the remaining defects is inessential, since each pair contains at
least one non-problematic defect, so we can distribute the cost of the latter
over the pair. In particular, if both are non-problematic, we will just add
the two separate contributions.

\subsubsection{Gluing process}

In this section we will construct for every layered event $\left(  \hat{\pi
}_{\tau}\right)  _{c}^{N}$ another layered event, $\phi_{2}\left(  \left(
\hat{\pi}_{\tau}\right)  _{c}^{N}\right)  ,$ of a similar type. The new event
will have less frustrated layers, and, what is most important, it will have no
problematic pairs of defects. More precisely, we prove the following:

\begin{lemma}
For any event $\left(  \hat{\pi}_{\tau}\right)  _{c}^{N}$ with $l$ problematic
pairs one can construct the event $\phi_{2}\left(  \left(  \hat{\pi}_{\tau
}\right)  _{c}^{N}\right)  ,$ such that:

\begin{enumerate}
\item
\[
\mu_{N,L}^{\beta,q}\left(  \left(  \hat{\pi}_{\tau}\right)  _{c}^{N}\right)
\leq a^{2lN^{2}}\mu_{N,L}^{\beta,q}\left(  \phi_{2}\left(  \left(  \hat{\pi
}_{\tau}\right)  _{c}^{N}\right)  \right)  ,
\]

\item all defects of $\phi_{2}\left(  \left(  \hat{\pi}_{\tau}\right)
_{c}^{N}\right)  $ can be paired in such a way that no pair is problematic,

\item the number of frustrated layers in $\phi_{2}\left(  \left(  \hat{\pi
}_{\tau}\right)  _{c}^{N}\right)  $ is $\left\Vert \hat{\pi}_{c}\right\Vert
-2l$.
\end{enumerate}
\end{lemma}

\begin{proof}
We proceed by induction on the number $l$ of problematic pairs, successively
removing every such pair and producing instead a factor $a^{2N^{2}}$.

We consider first the case when the two defects paired are problematic (or
e-problematic) signed defects $G_{i}$ and $G_{2k-i}$, with $1\leq i\leq k-1$.
We assume that the sign of $G_{i}$ (and therefore of $G_{2k-i}$) is minus; the
plus case is even simpler, since both defects are then non-exceptional
problematic defects.

We remind the reader that $G_{2k-i}$ consists of a sequence of 3 cubes: in
ascending order we first meet one pure disordered cube, then one frustrated,
followed by one pure ordered cube. All the bonds not in the ordered cube are
disordered. $G_{i}$ may be of problematic or e-problematic type, when
$i=1.$ In the first case it consists of $l_{i}=3$ cubes. In the second case it
will be convenient for us to include in the count of the cubes also the
\textquotedblleft virtual\textquotedblright\ disordered cube in the layer
$\left\{  -1\leq z\leq0\right\}  $, so we put $l_{1}$to be $3$, when the
e-defect has one frustrated and one ordered cube, and we put $l_{1}=4$ when
the e-problematic defect has two frustrated cubes plus one ordered on the top.
Note that in any case the first frustrated cube of the defect has at least 3
vertical disordered bonds. Each of $G_{j}$-s comes with the boundary condition
-- a configuration $\eta_{j}\in\Omega_{\partial G_{j}.}$

\textbf{The first step }of the gluing process is to make a global rotation,
$\Phi_{1},$ of the spin system in the slab $S_{i}=\left\{  a_{i}+2\leq z\leq
a_{2k-i}\right\}  $, so as to make the configuration $\eta_{i}^{t}$-- the
configuration on the plane $\left\{  z=b_{i}\right\}  ,$the top boundary
condition of the lower defect $G_{i}$-- to be closer to $\eta_{2k-i}^{t}$, the
top boundary condition of the defect $G_{2k-i}$. If the defect $G_{1}$happens
to be an e-problematic defect, then the slab $S_{1}=\left\{  1\leq z\leq
a_{2k-1}\right\}  $ by definition.

The configurations $\eta_{i}^{t}$ and $\eta_{2k-i}^{t}$ are two periodic
ordered configurations, defined by their restriction to any given plaquette,
so we write symbolically that $\eta_{i}^{t}=(s_{1},s_{2},s_{3},s_{4})$ and
$\eta_{2k-i}^{t}=(s_{1}^{\prime},s_{2}^{\prime},s_{3}^{\prime},s_{4}^{\prime
})$, where all $s_{1},...,s_{4}^{\prime}$ are just points of the discrete
circle $\mathbb{Z}_{q}.$ Since $\eta_{i}^{t}$ and $\eta_{2k-i}^{t}$ are
ordered, we can choose $s\in\left\{  s_{1},s_{2},s_{3},s_{4}\right\}  $ and
$s^{\prime}\in\left\{  s_{1}^{\prime},s_{2}^{\prime},s_{3}^{\prime}%
,s_{4}^{\prime}\right\}  $ such that for all $j=1,2,3,4$
\[
\left\vert s-s_{j}\right\vert \leq1,\text{ and }\left\vert s^{\prime}%
-s_{j}^{\prime}\right\vert \leq1.
\]
We will call the values $s,s^{\prime}$ the dominant values of the boundary
conditions. Now for every $\sigma\in\left(  \hat{\pi}_{\tau}\right)  _{c}^{N}$
we define $\Phi_{1}\left(  \sigma\right)  $ by
\[
\left[  \Phi_{1}\left(  \sigma\right)  \right]  \left(  x,y,z\right)
=\left\{
\begin{array}
[c]{cc}%
\sigma(x,y,z)+\left(  s-s^{\prime}\right)  & \text{ if }z\mathbf{\in}\left[
a_{i}+2,a_{2k-i}\right]  \mathbf{,}\\
\sigma(x,y,z) & \text{ otherwise.}%
\end{array}
\right.
\]
The transformation $\Phi_{1}$ is bijective.

The result thus achieved is that the configurations $\Phi_{1}\left(  \eta
_{i}^{t}\right)  $ and $\Phi_{1}\left(  \eta_{2k-i}^{t}\right)  $ are
relatively close to each other.

\textbf{The second (and the last) step} of the gluing process is to apply to
the system in the slab $S_{i}$ the reflection $\Phi_{2}$ in its middle
horizontal plane, thus bringing the upper part of $\Phi_{1}\left(
G_{i}\right)  $ in contact with $\Phi_{1}\left(  G_{2k-i}\right)  $:
\[
\left[  \Phi_{2}\left(  \sigma\right)  \right]  \left(  x,y,z\right)
=\left\{
\begin{array}
[c]{cc}%
\sigma\left(  x,y,a_{i}+a_{2k-i}+2-z\right)  & \text{ if }z\mathbf{\in}\left[
a_{i}+2,a_{2k-i}\right]  \mathbf{,}\\
\sigma\left(  x,y,z\right)  & \text{ otherwise.}%
\end{array}
\right.
\]
See Figure \ref{fig=gluing} for a sketch of this second step.
\begin{figure}
\centering
\includegraphics[width=0.9\textwidth]{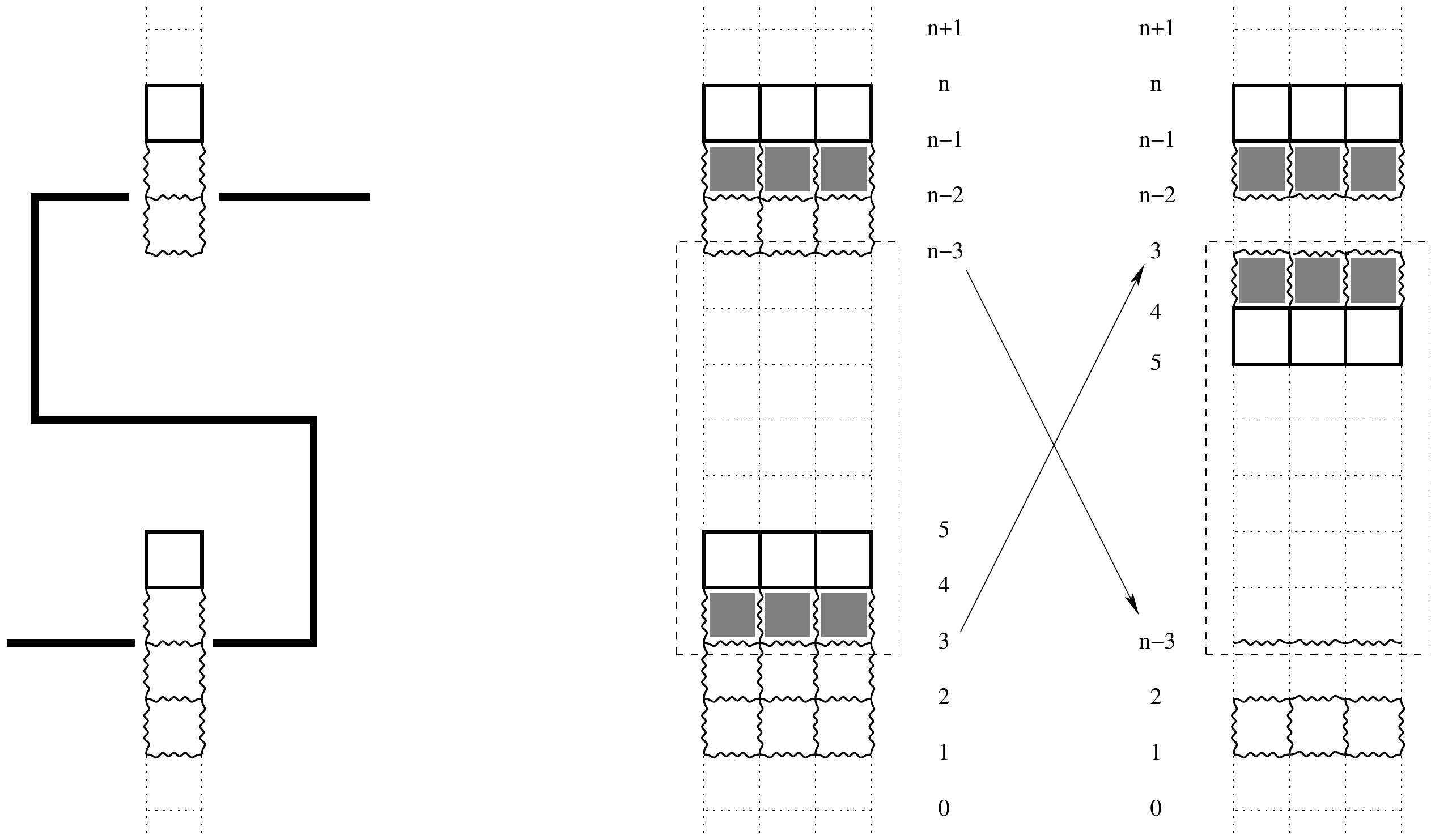}
\caption{Second step of the gluing transformation. From left to
right: two problematic defects (with minus signs), generated by a
piece of interface; the corresponding reflected event (made of two
defect sheets); the result of the gluing operation.}
\label{fig=gluing}
\end{figure}
(Again, for $G_{1}$ being e-problematic, the reflection is done in the slab
$S_{1}=\left\{  1\leq z\leq a_{2k-1}\right\}  ,$ with respect to the plane
$\left\{  z=\frac{1+a_{2k-1}}{2}+1\right\}  .$) The composition $\Phi=\Phi
_{2}\Phi_{1}$ is bijective. Note that for every configuration $\sigma
\in\left(  \hat{\pi}_{\tau}\right)  _{c}^{N}$ all the bonds connecting the
slab $S_{i}$ with its outside are disordered, except at most $\frac{N^{2}}{4}$
vertical bonds when $G_{1}$ is an e-problematic defect. Thus $\Phi$ can
increase the energy of the resulting configuration by $\frac{N^{2}}{4}$ units,
which is the possible number of ordered bonds turning into disordered ones
after the rotation: $H\left(  \sigma\right)  -H\left(  \Phi\left(
\sigma\right)  \right)  \geq-\frac{N^{2}}{4}$. Therefore we get %
\begin{equation}
\mu_{N,L}^{\beta,q}\left(  \left(  \hat{\pi}_{\tau}\right)  _{c}^{N}%
(\eta)\right)  \leq e^{\frac{\beta N^{2}}{4}}\mu_{N,L}^{\beta,q}\left(
\Phi\left[  \left(  \hat{\pi}_{\tau}\right)  _{c}^{N}(\eta)\right]  \right)  .
\label{rotate}%
\end{equation}

Let us describe the event $\Phi\left(  \left(  \hat{\pi}_{\tau}\right)
_{c}^{N}(\eta)\right)  $. Consider the images $\tilde{F}_{j}^{N}=\Phi\left(
F_{j}^{N}\right)  .$ If $F_{j}$ is between $G_{i}$ and $G_{2k-i}$, it is clear
that $\tilde{F}_{j}$ has exactly the same properties as $F_{j}$, up to shift
and reversal of pattern. Moreover, we will have $\tilde{\eta}_{j}^{t}%
=\Phi\left(  \eta_{j}^{b}\right)  $, $\tilde{\eta}_{j}^{b}=\Phi\left(
\eta_{j}^{t}\right)  $ as boundary conditions around $\tilde{F}_{j}$. If
$F_{j}$ is before $G_{i}$ of after $G_{2k-i}$, we have $\tilde{F}_{j}=F_{j}$
and $\tilde{\eta}_{j}=\eta_{j}$. The pattern $\tilde{\tau}=\Phi(\tau)$ is
defined in the following natural way: it coincides with $\tau$ outside the
slab $S_{i}$, and with a reflection of $\tau$ inside this slab.

We will now focus on what happened to $G_{i}^{N}$ and $G_{2k-i}^{N}$ ; we
denote by $\tilde{G}_{2k-i}^{N}$ the restriction of $\Phi\left(  \left(
\hat{\pi}_{\tau}\right)  _{c}^{N}\right)  $ to the slab
\[
\tilde{\Lambda}_{2k-i}=\left\{  a_{2k-i}-l_{i}+2\leq z\leq b_{2k-i}%
=a_{2k-i}+3\right\}  ,
\]
which is at most 5-cubes wide, since $l_{i}\leq4$.

If $\tilde{G}_{2k-i}^{N}$ occurs, we have two slabs -- $\left\{
a_{2k-i}-l_{i}+2\leq z\leq a_{2k-i}-l_{i}+3\right\}  $ and $\left\{
a_{2k-i}+2\leq z\leq a_{2k-i}+3\right\}  $ -- filled with ordered bonds, and
one slab -- $\left\{  a_{2k-i}+1\leq z\leq a_{2k-i}+2\right\}  $ -- filled
with disordered bonds; actually, the pattern of the bonds is fixed, except for
the $N^{2}$ vertical bonds of the slab $\left\{  a_{2k-i}\leq z\leq
a_{2k-i}+1\right\}  $. Since the boundary conditions $\tilde{\eta}%
_{2k-i}=\left(  \Phi\left(  \eta_{i}^{t}\right)  ,\eta_{2k-i}^{t}\right)  $
around this defect are very close to each other, we will be able to derive the
following estimate:
\begin{equation}
e^{\frac{\beta N^{2}}{4}}Z_{\tilde{\Lambda}_{2k-i}}^{\tilde{\eta}_{2k-i}%
}\left(  \tilde{G}_{2k-i}^{N}(\eta)\right)  \leq a^{2N^{2}}\cdot
Z_{\tilde{\Lambda}_{2k-i}}^{\tilde{\eta}_{2k-i}}, \label{32}%
\end{equation}
leading to
\begin{equation}
e^{\frac{\beta N^{2}}{4}}\mu_{N,L}^{\beta,q}\left(  \Phi\left(  \left(
\hat{\pi}_{\tau}\right)  _{c}^{N}(\eta)\right)  \right)  \leq a^{2N^{2}}%
\cdot\mu_{N,L}^{\beta,q}\left(  \left(  \tilde{\pi}_{\tilde{\tau}}\right)
_{c}^{N}(\tilde{\eta})\right)  , \label{33}%
\end{equation}
where $\left(  \tilde{\pi}_{\tilde{\tau}}\right)  _{c}^{N}(\tilde{\eta})$ is
the event that for all $j$ such that $F_{j}\notin\left\{  G_{i},G_{2k-i}%
\right\}  $, $\tilde{F}_{j}^{N}$ occurs, that the type on plaquettes of
$\partial^{o}\tilde{F}_{j}$ is given by $\tilde{\tau}=\Phi(\tau)$ and that at
the boundaries of $\tilde{F}_{j}^{N}$ the configuration agrees with
$\tilde{\eta}_{j}^{N}$.\medskip

The remaining case of a pair of problematic defects $F_{j},F_{k}$ with one of
them -- say, the upper one, $F_{k}$ -- having both end-cubes ordered, is even
simpler. Namely, it is enough to perform a global rotation in a suitable slab,
which will make the two (ordered) boundary conditions of the defect $F_{k}$
close enough, as it was the case in the first step above. After that, the
defect $F_{k}$ can be treated in precisely the same way as the defect
$\tilde{G}_{i}^{N}$ of the preceding paragraph was treated. To define the
rotation needed we take any slab $\left\{  d_{j}\leq z\leq d_{j}+1\right\}  $
inside the defect $F_{j}$, which has at least 3 disordered vertical bonds.
Such a slab clearly exists by definition. Then we do the rotation $\Phi$ of
all the spins in the slab $\left\{  d_{j}+1\leq z\leq a_{k}+2\right\}  $ by
the angle $s^{t}-s^{b},$ where $s^{t}$ and $s^{b}$ are the dominant values of
the boundary conditions $\eta_{k}^{t}$ and $\eta_{k}^{b}$ of the defect
$F_{k},$ leaving all other spins unchanged. Since $\Phi$ does not increase the
energy by more than $\frac{N^{2}}{4}$ units, we have reduced our case to the
one already considered.

Applying the above arguments to each problematic pair, we get rid of all of
them, getting a factor of $a^{2N^{2}}$ for each pair. We denote by $\phi_{2}$
the composition of the several transformations described above, which were
needed through the gluing process. Then $\phi_{2}\left(  F_{j}^{N}\right)  $
will be the family of remaining reflected defects, not yet treated, with
$\phi_{2}\left(  \eta_{c}\right)  $ being their boundary conditions. We denote
by $\phi_{2}\left(  \left(  \hat{\pi}_{\tau}\right)  _{c}^{N}\right)  $ the
event that all these defects occur and that the configuration takes the
prescribed values $\phi_{2}\left(  \eta_{c}\right)  $ on corresponding planes.
Summarizing, the lemma follows from $\left(  \ref{rotate},\ref{33}\right)  $,
the proof of $\left(  \ref{32}\right)  $ being deferred to the next section.
\end{proof}

\subsection{Estimating defects}

The estimates proceed differently for problematic and non-problematic defects.
We begin with the case of non-problematic ones.

\subsubsection{\label{NP} Non-problematic defects: Proof of $\left(
\ref{017}\right)  $}

Thanks to the previous analysis, the proof of our main theorem is reduced to
estimating a non-problematic defect. The analysis will be divided into three
cases, according to the nature of boundary conditions around the defect :
disordered, mixed, or ordered.

In the reflected defect $F_{i}^{N}$, we denote by $K_{i}$ the number of
chaotic sites, which are sites with 6 adjacent disordered bonds; notice that
$K_{i}~=k_{i}\frac{N^{2}}{4},$ with $k_{i}$ being an integer (or zero), due to
the periodic structure of $F_{i}^{N}.$ We denote by $D_{i}$ the number of
inner disordered bonds of $F_{i}^{N},$ (those of the configurations $\eta_{i}$
are not included). Let us consider the connected components of the graph made
by ordered bonds of $F_{i}^{N}.$ Some of these components are vertical
segments, not attached to the boundary; let $Q_{i}$ be their number. Again,
$Q_{i}=~q_{i}\frac{N^{2}}{4}$ with integer $q_{i}$. The number of other
connected components of this graph is at most $2L_{i}N$. Indeed, every such
component contains at least one full horizontal line (and there are $2L_{i}N$
such lines).

Note that the number of sites to which at least one ordered bond is attached
is $L_{i}N^{2}-K_{i}\leq2m_{i}N^{2},$ while the number of connected components
in this ordered bonds graph is at most $Q_{i}+2L_{i}N.$ We have therefore the
following simple universal upper bound:
\[
Z_{\Lambda_{i}}^{\eta_{i}}\left(  F_{i}^{N}\right)  \leq3^{2m_{i}N^{2}%
}q^{K_{i}+Q_{i}+2L_{i}N}e^{\beta\left(  \left(  3L_{i}+1\right)  N^{2}%
-D_{i}\right)  }.
\]
Indeed, let us pick a point in every connected component of the ordered bond
graph. Then the factor $q^{Q_{i}+2L_{i}N}$ estimates the number of possible
spin configurations $\varkappa$ on these sites, while $3^{2m_{i}N^{2}}$ is the
estimate on the number of configurations on the ordered bond graph, given
$\varkappa.$ (If the spin value at one end of the ordered bond is fixed, then
at the other end the spin can have $3$ different values, see $\left(
\ref{20}\right)  $.) The factor $q^{K_{i}}$ is the number of configurations on
chaotic sites. Finally, $\left(  3L_{i}+1\right)  N^{2}-D_{i}$ is the energy estimate.

We will use different lower bounds, depending on the boundary conditions and
the temperature. They use some (heavy) combinatorics of the defects. We
postpone the proof of the relevant combinatorial statements till the end of
the paper.

\paragraph{Order--disorder.}

In this subsection we consider non-problematic defects with ordered boundary
condition at one end of the defect and disordered boundary condition at the
other. We have the bound
\[
Z_{\Lambda_{i}}^{\eta_{i}}\geq\left(  q-18\right)  ^{L_{i}N^{2}}+e^{3\beta
L_{i}N^{2}};
\]
here the first term estimates the partition function taken over fully
disordered configurations, while the second one -- the partition function
taken over fully ordered configurations. (In fact, it is enough to take just
one ordered configuration.) If $e^{\beta}\leq q^{1/3}$, we have
(omitting unimportant terms, not depending on $q$):
\begin{align*}
\frac{Z_{\Lambda_{i}}^{\eta_{i}}\left(  F_{i}^{N}\right)  }{Z_{\Lambda_{i}%
}^{\eta_{i}}} &  \leq9^{m_{i}N^{2}}\frac{q^{K_{i}+Q_{i}+2L_{i}N}%
e^{\beta\left(  \left(  3L_{i}+1\right)  N^{2}-D_{i}\right)  }}{\left(
q-18\right)  ^{L_{i}N^{2}}}\\
&  \leq9^{m_{i}N^{2}}\left(  \frac{1}{q}\right)  ^{\left(  D_{i}-N^{2}\right)
/3-K_{i}-Q_{i}-2L_{i}N}.
\end{align*}
If $e^{\beta}\geq q^{1/3}$,
\begin{align*}
\frac{Z_{\Lambda_{i}}^{\eta_{i}}\left(  F_{i}^{N}\right)  }{Z_{\Lambda_{i}%
}^{\eta_{i}}} &  \leq9^{m_{i}N^{2}}\frac{q^{K_{i}+Q_{i}+2L_{i}N}%
e^{\beta\left(  \left(  3L_{i}+1\right)  N^{2}-D_{i}\right)  }}{e^{3\beta
L_{i}N^{2}}}\\
&  \leq9^{m_{i}N^{2}}\left(  \frac{1}{q}\right)  ^{\left(  D_{i}-N^{2}\right)
/3-K_{i}-Q_{i}-2L_{i}N}%
\end{align*}
By $\left(  \ref{63}\right)  $ below we can take $\alpha^{\prime}>0$ such
that
\[
\left(  D_{i}-N^{2}\right)  /3-K_{i}-Q_{i}\geq2\alpha^{\prime}m_{i}N^{2}.
\]
Since $L_{i}\leq2m_{i}$, for all $N\ $large enough and all order--disorder
defects $F_{i}$,
\[
\left(  D_{i}-N^{2}\right)  /3-K_{i}-Q_{i}-2L_{i}N\geq\alpha^{\prime}%
m_{i}N^{2}.
\]
Therefore, for all $\beta$ and all such defects,
\[
\frac{Z_{\Lambda_{i}}^{\eta_{i}}\left(  F_{i}^{N}\right)  }{Z_{\Lambda_{i}%
}^{\eta_{i}}}\leq9^{m_{i}N^{2}}q^{-\alpha^{\prime}m_{i}N^{2}},
\]
and the desired estimate is valid with $a(q)=9q^{-\alpha^{\prime}}$.

\paragraph{Order--Order.}

As in the order--disorder case, we have%

\[
Z_{\Lambda_{i}}^{\eta_{i}}\geq\left(  q-18\right)  ^{L_{i}N^{2}}+e^{3\beta
L_{i}N^{2}},
\]
so we will be done by the previous analysis, if the estimate
\[
\left(  D_{i}-N^{2}\right)  /3-K_{i}-Q_{i}\geq2\alpha^{\prime}m_{i}N^{2}%
\]
still holds for the order-order case. This is indeed so, see again $\left(
\ref{63}\right)  .$ Therefore for all $\beta>0$
\[
\frac{Z_{\Lambda_{i}}^{\eta_{i}}\left(  F_{i}^{N}\right)  }{Z_{\Lambda_{i}%
}^{\eta_{i}}}\leq9^{m_{i}N^{2}}q^{-\alpha^{\prime}m_{i}N^{2}}.
\]

\paragraph{Bulk Disorder--Disorder.}

We have
\[
Z_{\Lambda_{i}}^{\eta_{i}}\geq\left(  q-18 \right)  ^{L_{i}N^{2} }+
e^{\beta\left(  3L_{i}-1\right)  N^{2}}.
\]

If $e^{\beta}\leq q^{L_{i}/(3L_{i}-1)}$, we have (omitting unimportant terms,
not depending on $q$) :
\begin{align*}
\frac{Z_{\Lambda_{i}}^{\eta_{i}}\left(  F_{i}^{N}\right)  }{Z_{\Lambda_{i}%
}^{\eta_{i}}}  &  \leq9^{m_{i}N^{2}}\frac{q^{K_{i}+Q_{i}+2L_{i}N}%
e^{\beta\left(  \left(  3L_{i}+1\right)  N^{2}-D_{i}\right)  }}{\left(
q-18\right)  ^{L_{i}N^{2}}}\\
&  \leq9^{m_{i}N^{2}}\left(  \frac{1}{q}\right)  ^{\frac{L_{i}}{3L_{i}%
-1}\left[  D_{i}-2N^{2}-\frac{3L_{i}-1}{L_{i}}\left(  K_{i}+Q_{i}\right)
\right]  -2L_{i}N}.
\end{align*}
If $e^{\beta}\geq q^{L_{i}/(3L_{i}-1)}$,
\begin{align*}
\frac{Z_{\Lambda_{i}}^{\eta_{i}}\left(  F_{i}^{N}\right)  }{Z_{\Lambda_{i}%
}^{\eta_{i}}}  &  \leq9^{m_{i}N^{2}}\frac{q^{K_{i}+Q_{i}+2L_{i}N}%
e^{\beta\left(  \left(  3L_{i}+1\right)  N^{2}-D_{i}\right)  }}{e^{\beta
\left(  3L_{i}-1\right)  N^{2}}}\\
&  \leq9^{m_{i}N^{2}}\left(  \frac{1}{q}\right)  ^{\frac{L_{i}}{3L_{i}%
-1}\left[  D_{i}-2N^{2}-\frac{3L_{i}-1}{L_{i}}\left(  K_{i}+Q_{i}\right)
\right]  -2L_{i}N}.
\end{align*}
By $\left(  \ref{70}\right)  $ we can take $\alpha^{\prime}>0$ such that
\[
D_{i}-2N^{2}-\frac{3L_{i}-1}{L_{i}}\left(  K_{i}+Q_{i}\right)  \geq
6\alpha^{\prime}m_{i}N^{2}.
\]
Since $L_{i}\leq2m_{i}$, for all $N\ $large enough and all disorder--disorder
defects $F_{i}$,
\[
\frac{L_{i}}{3L_{i}-1}\left[  D_{i}-2N^{2}-\frac{3L_{i}-1}{L_{i}}\left(
K_{i}+Q_{i}\right)  \right]  -2L_{i}N\geq\alpha^{\prime}m_{i}N^{2}.
\]
Therefore, for all $\beta$ and all such defects,
\[
\frac{Z_{\Lambda_{i}}^{\eta_{i}}\left(  F_{i}^{N}\right)  }{Z_{\Lambda_{i}%
}^{\eta_{i}}}\leq9^{m_{i}N^{2}}q^{-\alpha^{\prime}m_{i}N^{2}},
\]
and the desired estimate is valid with $a(q)=9q^{-\alpha^{\prime}}$.

\paragraph{Boundary Disorder--Disorder.}

We have
\[
Z_{\Lambda_{i}}^{\eta_{i}}\geq\left(  q-18\right)  ^{L_{i}N^{2}}%
+e^{\beta\left(  3L_{i}-\frac{3}{4}\right)  N^{2}}.
\]
(We have $\frac{3}{4}$ in the energy estimate $\left(  3L_{i}-\frac{3}%
{4}\right)  N^{2}$ due to the fact that at least one quarter of the boundary
bonds will be ordered.)

If $e^{\beta}\leq q^{L_{i}/\left(  3L_{i}-\frac{3}{4}\right)  }$,
\begin{align*}
\frac{Z_{\Lambda_{i}}^{\eta_{i}}\left(  F_{i}^{N}\right)  }{Z_{\Lambda_{i}%
}^{\eta_{i}}}  &  \leq9^{m_{i}N^{2}}\frac{q^{K_{i}+Q_{i}+2L_{i}N}%
e^{\beta\left(  \left(  3L_{i}+1\right)  N^{2}-D_{i}\right)  }}{\left(
q-18\right)  ^{L_{i}N^{2}}}\\
&  \leq9^{m_{i}N^{2}}\left(  \frac{1}{q}\right)  ^{\frac{L_{i}}{6\left(
L_{i}-\frac{1}{4}\right)  }\left[  2D_{i}-\frac{7}{2}N^{2}-\frac{6\left(
L_{i}-\frac{1}{4}\right)  }{L_{i}}\left(  K_{i}+Q_{i}\right)  \right]
-2L_{i}N}.
\end{align*}
If $e^{\beta}\geq q^{L_{i}/\left(  3L_{i}-\frac{3}{4}\right)  }$,
\begin{align*}
\frac{Z_{\Lambda_{i}}^{\eta_{i}}\left(  F_{i}^{N}\right)  }{Z_{\Lambda_{i}%
}^{\eta_{i}}}  &  \leq9^{m_{i}N^{2}}\frac{q^{K_{i}+Q_{i}+2L_{i}N}%
e^{\beta\left(  \left(  3L_{i}+1\right)  N^{2}-D_{i}\right)  }}{e^{\beta
\left(  3L_{i}-\frac{3}{4}\right)  N^{2}}}\\
&  \leq9^{m_{i}N^{2}}\left(  \frac{1}{q}\right)  ^{\frac{L_{i}}{6\left(
L_{i}-\frac{1}{4}\right)  }\left[  2D_{i}-\frac{7}{2}N^{2}-\frac{6\left(
L_{i}-\frac{1}{4}\right)  }{L_{i}}\left(  K_{i}+Q_{i}\right)  \right]
-2L_{i}N}.
\end{align*}
Below in $\left(  \ref{70bis}\right)  $ we will show that for some
$\alpha^{\prime}>0$
\[
2D_{i}-\frac{7N^{2}}{2}-\frac{6\left(  L_{i}-\frac{1}{4}\right)  }{L_{i}%
}(K_{i}+Q_{i})\geq12\alpha^{\prime}m_{i}N^{2}.
\]
Since $L_{i}\leq2m_{i}$, for all $N$ large enough and all disorder-disorder
boundary defects $F_{i}$
\[
\frac{L_{i}}{6\left(  L_{i}-\frac{1}{4}\right)  }\left[  2D_{i}-\frac{7}%
{2}N^{2}-\frac{6\left(  L_{i}-\frac{1}{4}\right)  }{L_{i}}(K_{i}%
+Q_{i})\right]  -2L_{i}N\geq\alpha^{\prime}m_{i}N^{2}.
\]
Therefore for all $\beta$ and all such defects
\[
\frac{Z_{\Lambda_{i}}^{\eta_{i}}\left(  F_{i}^{N}\right)  }{Z_{\Lambda_{i}%
}^{\eta_{i}}}\leq9^{m_{i}N^{2}}q^{-\alpha^{\prime}m_{i}N^{2}},
\]
and the desired estimate is valid with $a(q)=9q^{-\alpha^{\prime}}$.
$\blacksquare$

\subsubsection{Glued pair of problematic defects: Proof of $\left(
\ref{32},\ref{33}\right)  $}

We will analyze the defect $\tilde{G}_{i}^{N},$ generated by the gluing
process, and will prove the estimates $\left(  \ref{32},\ref{33}\right)  $.
The defect $\tilde{G}_{i}^{N}$ is at most $5$-cubes wide, both end-layers are
ordered, and all vertical bonds attached to the top cube are disordered; we
notice that some vertical bonds in the third layer from the top may be
ordered, possibly in a non-periodic way. We fix the pattern $V$ of these extra
vertical ordered bonds, $\tilde{G}_{i}^{N}\left(  V\right)  $ denoting the
restriction of $\tilde{G}_{i}^{N}$ to configurations agreeing with the pattern
$V$. We will now estimate the partition function $Z_{\tilde{\Lambda}_{i}%
}^{\tilde{\eta}_{i}}\left(  \tilde{G}_{i}^{N}\left(  V\right)  \right)  $ in
its slab $\tilde{\Lambda}_{i}=\left\{  \tilde{a}_{i}\leq z\leq\tilde{b}%
_{i}\right\}  ,$ and write $\tilde{L}_{i}=\tilde{b}_{i}-\tilde{a}%
_{i}-1\leq4$.

The number of configurations in the slab $\tilde{\Lambda}_{i},$ such that the
event $\tilde{G}_{i}^{N}\left(  V\right)  $ occurs, is bounded from above by
$3^{\tilde{L}_{i}N^{2}}q^{K+Q+2\tilde{L}_{i}N}$, where $K,Q$ depend on $V$;
every such configuration $\sigma^{(i)}$ has energy $H^{(i)}(\sigma
^{(i)})=D-\left(  3\tilde{L}_{i}+1\right)  N^{2}$, where $D$ also depends on
$V$. Combining this we get:
\[
Z_{\tilde{\Lambda}_{i}}^{\tilde{\eta}_{i}}\left(  \tilde{G}_{i}^{N}\left(
V\right)  \right)  \leq3^{\tilde{L}_{i}N^{2}}q^{K+Q+2\tilde{L}_{i}N}%
e^{\beta\left(  (3\tilde{L}_{i}+1)N^{2}-D\right)  }.
\]
Now we need a lower bound on $Z_{\tilde{\Lambda}_{i}}^{\tilde{\eta}_{i}}.$ We
will use one consisting of two contributions: the first is obtained by summing
over high temperature configurations, while the second -- by summing over low
temperature ones.

For high temperatures, we just integrate over configurations with zero energy,
the set of such configurations containing at least $\left(  q-18\right)
^{\tilde{L}_{i}N^{2}}$ configurations.

For low temperatures, we simply take one single configuration with minimal
energy under given boundary conditions. Let us check that this minimum equals
to $-(3\tilde{L}_{i}+1)N^{2}$. Indeed, since the (periodic) configurations
$\left(  \tilde{\eta}_{i}\right)  ^{t}$ on $\left\{  z=\tilde{b}_{i}\right\}
$ and $\left(  \tilde{\eta}_{i}\right)  ^{b}$ on $\left\{  z=\tilde{a}%
_{i}\right\}  $ have by construction the common dominant value, $s$, the
constant configuration $\sigma_{s}\equiv s$ in $\left\{  \tilde{a}_{i}+1\leq
z\leq\tilde{b}_{i}-1\right\}  $ -- the interior of $\tilde{\Lambda}_{i}$,
taken with boundary conditions $\tilde{\eta}_{i}$, has all bonds in
$\tilde{\Lambda}_{i}$ ordered.

Gathering all this we have: %
\[
Z_{\tilde{\Lambda}_{i}}^{\tilde{\eta}_{i}}\geq\left(  q-18\right)  ^{\tilde
{L}_{i}N^{2}}+e^{(3\tilde{L}_{i}+1)N^{2}\beta}.
\]
 If $e^{\beta}\leq q^{\tilde{L}_{i}/(3\tilde{L}_{i}+1)}$, we use
$Z_{\tilde{\Lambda}_{i}}^{\tilde{\eta}_{i}}\geq\left(  q-18\right)
^{\tilde{L}_{i}N^{2}}$ to get
\begin{equation}
e^{\frac{1}{4}\beta N^{2}}\cdot\frac{Z_{\tilde{\Lambda}_{i}}^{\tilde{\eta}%
_{i}}\left(  \tilde{G}_{i}^{N}\left(  V\right)  \right)  }{Z_{\tilde{\Lambda
}_{i}}^{\tilde{\eta}_{i}}}\leq\left(  \frac{3q}{q-18}\right)  ^{\tilde{L}%
_{i}N^{2}}\left(  q^{-\frac{\tilde{L}_{i}}{3\tilde{L}_{i}+1}}\right)
^{D-\frac{1}{4}N^{2}-\frac{3\tilde{L}_{i}+1}{\tilde{L}_{i}}(K+Q+2\tilde{L}%
_{i}N)} \label{glued_pair_high}%
\end{equation}
If $e^{\beta}\geq q^{\tilde{L}_{i}/(3\tilde{L}_{i}+1)}$, we use $Z_{\tilde
{\Lambda}_{i}}^{\tilde{\eta}_{i}}\geq e^{(3\tilde{L}_{i}+1)\beta N^{2}}$ to get%

\begin{equation}
\label{glued_pair_low}e^{\frac{1}{4} \beta N^{2}} \cdot\frac{Z_{\tilde
{\Lambda}_{i}}^{\tilde{\eta}_{i}}\left(  \tilde{G}_{i} ^{N}\left(  V\right)
\right)  }{Z_{\tilde{\Lambda}_{i}}^{\tilde{\eta}_{i}} }\leq3^{\tilde{L}%
_{i}N^{2}} \left(  q^{-\frac{\tilde{L}_{i}}{3\tilde{L}_{i}+1}} \right)
^{D-\frac{1}{4}N^{2}-\frac{3\tilde{L}_{i}+1}{\tilde{L}_{i}}(K+Q+2\tilde{L}%
_{i}N)} .
\end{equation}
We will use the following

\begin{lemma}
\label{glued_pair_estimate} For any pattern $V$ of ordered bonds in the third
layer from the top, and all $N$ large enough
\begin{equation}
D-\frac{3\tilde{L}_{i}+1}{\tilde{L}_{i}}(K+Q+2\tilde{L}_{i}N)\geq\left(
\frac{1}{4}+\frac{1}{5}\right)  N^{2}.
\end{equation}

\end{lemma}

\begin{proof}
We recall that the ordered cubes at end-points of the defect are always
disconnected in the ordered graph corresponding to $V$, because of the
vertical disordered bonds in the second layer from the top; using Lemma
\ref{disconnected} below we get $D-3(K+Q)\geq N^{2}$. Moreover, it is clear
that $K+Q\leq(\tilde{L}_{i}-2)N^{2}$ and thus $D-\frac{3\tilde{L}_{i}%
+1}{\tilde{L}_{i}}(K+Q)\geq N^{2}-\frac{\left(  \tilde{L}_{i}-2\right)
}{\tilde{L}_{i}}N^{2}=\frac{2}{\tilde{L}_{i}}N^{2}$. Our lemma now follows
from $\tilde{L}_{i}\leq4$ since $\frac{2}{\tilde{L}_{i}}N^{2}\geq\frac{1}%
{2}N^{2}\geq\frac{1}{4}N^{2}+\frac{1}{5}N^{2}+2\tilde{L}_{i}N$, provided $N$
is large enough.
\end{proof}

Since there are $2^{N^{2}}$ possible patterns $V$, Lemma
\ref{glued_pair_estimate} together with $\left(  \ref{glued_pair_high}%
,\ref{glued_pair_low}\right)  $ give for all $\beta$:
\[
e^{\frac{1}{4}\beta N^{2}}\cdot\frac{Z_{\tilde{\Lambda}_{i}}^{\tilde{\eta}%
_{i}}\left(  \tilde{G}_{i}^{N}\right)  }{Z_{\tilde{\Lambda}_{i}}^{\tilde{\eta
}_{i}}}\leq2^{N^{2}}\left(  \left(  \frac{3q}{q-18}\right)  ^{\tilde{L}_{i}%
}q^{-\frac{1}{5}\cdot\frac{\tilde{L}_{i}}{3\tilde{L}_{i}+1}}\right)  ^{N^{2}%
}\leq a^{2N^{2}},
\]
with $a(q)=\sqrt{2\left(  \frac{3q}{q-18}\right)  ^{4}q^{-3/50}}$, since
$\tilde{L}_{i}\leq4$ and also $\frac{\tilde{L}_{i}}{3\tilde{L}_{i}+1}\geq
\frac{3}{10}$. $\blacksquare$

\section{Combinatorial estimates for bulk defects}

\label{app:combinatorics}

We prove here the needed combinatorial estimates on \textit{non-problematic}
defects, restricting the proof to defects in the bulk of the system (i.e. when
the defect is not stuck to the bottom boundary), and divide this proof into
three parts according to the nature of the boundary conditions around the
defect. We introduce the number $d,$ which equals the number of disordered
cubes at the ends of our defect, i.e.
\[
d=\left\{
\begin{array}
[c]{cc}%
0 & \text{for the order-order bc,}\\
1 & \text{for the order-disorder bc,}\\
2 & \text{for the disorder-disorder bc.}%
\end{array}
\right.
\]
The case of boundary defects is more involved and is deferred to the next section.

For $d=0$ or $d=1$ non-problematic bulk defect $F$ with $m\geq1$ frustrated
cubes, and its reflection $F^{N}$ we will prove the relation
\begin{equation}
2D-6K-6Q\geq2N^{2}+\alpha mN^{2} \label{63}%
\end{equation}
for some universal $\alpha>0,$ where $D$, $K$ and $Q$ are the characteristics
of $F^{N},$ introduced above. For $d=2$ non-problematic bulk defect $F$ with
$m\geq1$ frustrated cubes we will prove
\begin{equation}
2D-2\frac{\left(  3L-1\right)  }{L}(K+Q)\geq4N^{2}+\alpha mN^{2}. \label{70}%
\end{equation}

We introduce the set $\mathrm{K}$ of chaotic sites and the set $\mathrm{D}$ of
disordered bonds in $F,$ $\left\vert \mathrm{K}\right\vert =K,$ $\left\vert
\mathrm{D}\right\vert =D,$ and we rewrite $6K$ as a double sum
\[
6K=\sum_{x\in\mathrm{K}}\sum_{e:x\in e}\mathbb{I}_{e\in\mathrm{D}},
\]
to get
\[
2D=6K+dN^{2}+\left\vert \partial^{1}O\right\vert +2\left\vert \partial
^{2}O\right\vert ,
\]
where $O$ is the graph of ordered bonds in $F^{N}$, and $\partial^{n}O$
denotes the set of disordered bonds with $n$ vertices belonging to $O,$
$n=1,2$; the term $dN^{2}$ comes from the $dN^{2}$ vertical disordered bonds
in the boundary chaotic cubes (this is precisely where we use the fact that
the defect is in the bulk). We rewrite it as
\begin{equation}
2D=6K+dN^{2}+\sum_{j}\left\vert \partial X_{j}\right\vert +\sum_{j}\left\vert
\partial^{2}X_{j}\right\vert , \label{cr}%
\end{equation}
where $X_{j}$-s are the connected components of the ordered-bond graph of
$F^{N}$, $\partial X_{j}$ is the set of disordered bonds touching $X_{j}$, and
$\partial^{2}X_{j}$ the set of disordered bonds with both vertex in $X_{j}$.
When the number $m$ of frustrated cubes in the defect is small, we will use
for the derivation of $\left(  \ref{63}\right)  $ the above relation $\left(
\ref{cr}\right)  $ directly. For large $m$-s we will utilize its corollary,
which we will derive now.

\begin{lemma}
The relation $\left(  \ref{cr}\right)  $ implies that
\begin{equation}
2D-6K-6Q\geq dN^{2}+\frac{1}{2}mN^{2}+2Q. \label{62}%
\end{equation}

\end{lemma}

\begin{proof}
If $X_{j}$ is a vertical segment, not touching the boundary, we have
$\left\vert \partial X_{j}\right\vert =n_{j}+6$, where $n_{j}$ is the number
of frustrated cubes sharing a bond with $X_{j}$; also, $\partial^{2}X_{j}=0$.
Let us denote the set of these $j$-s by $J.$ For other components we use the
estimate:
\begin{equation}
\left\vert \partial X_{j}\right\vert +\left\vert \partial^{2}X_{j}\right\vert
\geq\frac{1}{2}n_{j}. \label{61}%
\end{equation}
To see it to hold, we first note that
\begin{equation}
\left\vert \partial X_{j}\right\vert +\left\vert \partial^{2}X_{j}\right\vert
\geq\frac{1}{4}\sum_{c}\left(  \left\vert \partial X_{j}\cap c\right\vert
+\left\vert \partial^{2}X_{j}\cap c\right\vert \right)  , \label{77}%
\end{equation}
where the summation goes over all cubes $c,$ contributing to $n_{j};$ we have
the factor $\frac{1}{4}$ due to the fact that every bond belongs to at most 4
cubes. We claim now that for every cube $c$ we have $\left\vert \partial
X_{j}\cap c\right\vert +\left\vert \partial^{2}X_{j}\cap c\right\vert \geq2.$
Indeed, either $c$ has at least two bonds from $\partial X_{j}$, or just one
such bond, $f.$ In the latter case, all other (eleven) bonds of $c$ belong to
$X_{j},$ and therefore $f$ belongs not only to $\partial X_{j},$ but also to
$\partial^{2}X_{j}.$ That proves $\left(  \ref{61}\right)  .$ Gathering all
this leads to:
\begin{align*}
2D-6K  &  \geq dN^{2}+6Q+\sum_{j\in J}n_{j}+\frac{1}{2}\sum_{j\notin J}n_{j}\\
&  \geq dN^{2}+8Q+\frac{1}{2}\sum_{j}n_{j},
\end{align*}
since for every $j\in J$ we have $\frac{1}{2}n_{j}\geq2.$ Finally, every
frustrated cube in $F^{N}$ contributes to at least one $n_{j},$ so we arrive
to%
\[
2D-6K-6Q\geq dN^{2}+\frac{1}{2}mN^{2}+2Q.
\]

\end{proof}

\subsection{Order--disorder ($d=1$): Proof of $\left(  \ref{63}\right)  $}

Here we consider a non-problematic defect with an ordered (disordered) cube at
the top (bottom).

\begin{enumerate}
\item If $m\geq3$, $\left(  \ref{62}\right)  $ gives
\[
2D-6(K+Q)-2N^{2}\geq\left(  \frac{m}{2}-1\right)  N^{2}\geq\frac{m}{6}N^{2},
\]
which is what we need.

\item Assume $m\leq2$. Let us consider the ordered connected component
$X_{0},$ containing the upper ordered cube; its boundary $\partial X_{0}$ has
at least $N^{2}$ bonds, with equality if and only if $X_{0}$ is the result of
multiple reflections of the upper ordered cube. Thus $\left(  \ref{cr}\right)
$ shows that $2D-6(K+Q)\geq2N^{2},$ with equality if and only if $Q=0$,
$\left\vert \partial X_{0}\right\vert =N^{2}$, and the set $\left\{
X_{j}\right\}  $ consists of only one component -- $X_{0}.$ The equality
therefore can occur only if the defect is problematic. Hence in the case
considered $2D-6(K+Q)\geq2N^{2}+\frac{N^{2}}{2}$.
\end{enumerate}

$\blacksquare$

\subsection{ Disorder--disorder ($d=2$): Proof of $\left(  \ref{70}\right)  $}

Here we consider a non-problematic reflected defect $F^{N}$ surrounded by two
chaotic layers. We want to obtain the bound $\left(  \ref{70}\right)  .$ In
fact, for most defects the stronger statement holds:
\begin{equation}
2D-6(K+Q)\geq4N^{2}+\alpha mN^{2}. \label{71}%
\end{equation}
Indeed, the relation $\left(  \ref{62}\right)  $ reads
\begin{equation}
2D-6K-6Q\geq2N^{2}+\frac{1}{2}mN^{2}+2Q, \label{73}%
\end{equation}
so the estimate $\left(  \ref{71}\right)  $ holds once $m\geq5$. So we assume
in the following that $m\leq4$; if $K=Q=0$, the simple fact that $D\geq6N^{2}$
is enough to get (\ref{71}), so we assume it is not the case. But then, it is
enough to show that $2D-6(K+Q)\geq4N^{2}$; indeed, (\ref{70}) will follow from
$L\leq2m=8$ and $K+Q\geq\frac{N^{2}}{4}$.

Next we note the following simple

\begin{lemma}
For any bulk defect with disorder--disorder b.c. ($d=2$), the existence of an
ordered horizontal bond $e$ implies
\[
\sum_{i\notin J}\left\vert \partial X_{i}\right\vert \geq2N^{2}.
\]

\end{lemma}

\begin{proof}
Indeed, in its column $c$ the bond $e$ has two horizontal adjacent bonds
$e^{\prime},e^{\prime\prime}$. If both of them are disordered, their
reflections produce $N^{2}$ horizontal bonds belonging to $\cup_{i\notin
J}\partial X_{i},$ while the reflections of the bond $e$ contain $\frac{N^{2}%
}{2}$ sites, each of which has a disordered bond from $\cup\partial X_{i}$
above it and another one below it. If $e^{\prime}$ is ordered and
$e^{\prime\prime}$ is disordered, we get similarly $\frac{N^{2}}{2}$
horizontal bonds and $\frac{3N^{2}}{2}$ vertical bonds in the boundaries. If
both $e^{\prime}$ and $e^{\prime\prime}$ are ordered, we get $2N^{2}$ vertical
bonds in the boundaries.
\end{proof}

The previous lemma, combined with $\left(  \ref{cr}\right)  ,$ reduces the
analysis to the case where there is no ordered horizontal bonds. In this last
case we have $2D-6K=2N^{2}+6Q+\sum_{j\in J}n_{j}$, where $n_{j}$ is the number
of frustrated cubes sharing a bond with $X_{j}.$ From this, we get
$2D-6(K+Q)\geq2N^{2}+mN^{2}\geq4N^{2}$, if $m\geq2$. If $m=1$, then $L=2$,
$K=2N^{2}-2Q$, $D=7N^{2}-Q$ with $Q\geq\frac{N^{2}}{4}$, so that
$2D-2\frac{3L-1}{L}(K+Q)=2D-5(K+Q)=4N^{2}+3Q\geq4N^{2}+\frac{3N^{2}}{4}$.

$\blacksquare$

\subsection{Order--order ($d=0$): Proof of $\left(  \ref{63}\right)  $}

We consider the reflection $F^{N}$ of a non-problematic defect $F$ with $m$
frustrated cubes, surrounded by two ordered cubes. Since every defect by
definition contains a disordered plaquette, every order-order defect has
$m\geq2$. We want to establish the relation $\left(  \ref{63}\right)  :\ $
$2D-6(K+Q)\geq2N^{2}+\alpha mN^{2}.$ For $m\geq5$ it follows immediately from
$\left(  \ref{62}\right)  $, so we assume that $m\leq4$. In this case the
relation $\left(  \ref{63}\right)  $ follows from the following two lemmas:

\begin{lemma}
\label{disconnected} For all defects with order--order b.c. and such that the
ordered cubes at the ends of the defect are disconnected in the ordered
graph,
\[
2D-6(K+Q)\geq2N^{2},
\]
with equality if and only if it is problematic.
\end{lemma}

\begin{proof}
We denote by $X_{0}$ and $X_{1}$ the two connected components corresponding to
the extreme ordered cubes. Then $\left\vert \partial X_{0}\right\vert
+\left\vert \partial X_{1}\right\vert \geq2N^{2}$ and $\left(  \ref{cr}%
\right)  $ shows the inequality, and we see that the case of equality is
precisely the problematic defect.
\end{proof}

\begin{lemma}
For all defects with order--order b.c. such that ordered cubes at the ends of
the defect belong to the same component,
\[
2D-6(K+Q) \geq3N^{2}.
\]

\end{lemma}

\begin{proof}
We denote by $X_{0}$ the component containing both ordered cubes. Our
assumption means that our defect contains a vertical disordered plaquette $P$.
Indeed, all the blobs defining our defect have only vertical plaquettes, since
the defect does not contain disordered cubes.

Looking at the two vertical lines passing through $P$, we see that each of
them is either completely ordered outside $P$, its unique disordered bond then
belonging to $\partial^{2}X_{0}$, or else it has two bonds in $\partial X_{0}%
$; therefore, the contribution of vertical bonds to $\left\vert \partial
X_{0}\right\vert +\left\vert \partial^{2}X_{0}\right\vert $ is at least
$2\frac{N^{2}}{2}=N^{2}.$

We shall now prove that the horizontal contribution to $\left\vert \partial
X_{0}\right\vert +\left\vert \partial^{2}X_{0}\right\vert $ is at least
$2N^{2}$. Let us look at the horizontal plaquette $P^{\prime},$ which contains
the bottom horizontal bond of $P.$ Of course, this bond is disordered. If some
ordered bonds of $P^{\prime}$ belong to $X_{0},$ then $\left\vert \partial
X_{0}\cap P^{\prime}\right\vert +\left\vert \partial^{2}X_{0}\cap P^{\prime
}\right\vert \geq2,$ as a simple counting shows. Otherwise, since there is an
ordered path through the defect, there is a vertical bond in $X_{0}$ touching
$P^{\prime}$ at a vertex $x$. By assumption, the two bonds of $P^{\prime}$
containing $x$ are disordered (since otherwise they would belong to $X_{0}$),
so they both are in $\partial X_{0}$. The same holds for the horizontal
plaquette $P^{\prime},$ which shares the top horizontal bond with $P,$ which
proves our claim.
\end{proof}

$\blacksquare$

\section{Combinatorial estimates for boundary defects}

Now we deal with the case when the defect is stuck to the bottom of the box.

\subsection{Order--disorder: Proof of $\left(  \ref{63}\right)  $}

Let us denote by $D^{b}$ the number of vertical disordered bonds attached to
the bottom boundary of $F^{N}$ and replacing in $\left(  \ref{62}\right)  $
the term $dN^{2}$ by $D^{b},$ we have the analog of $\left(  \ref{cr}\right)
$
\begin{equation}
2D=6K+D^{b}+\sum_{j}\left\vert \partial X_{j}\right\vert +\sum_{j}\left\vert
\partial^{2}X_{j}\right\vert , \label{78}%
\end{equation}
and the analog of $\left(  \ref{62}\right)  :$
\begin{equation}
2D-6K-6Q\geq D^{b}+\frac{1}{2}mN^{2}+2Q. \label{75}%
\end{equation}
(We remark for clarity that here $Q$ is the number of ordered vertical
segments, not touching both boundaries of the defect $F^{N}.$)

We recall the reader that we aim to prove the relation $\left(  \ref{63}%
\right)  $ for non-e-problematic defects. Note that the only $m=1$ boundary
defect with order--disorder b.c. is e-problematic. So in what follows we
assume that $m\geq2$.

If $m\geq5$, the relation $\left(  \ref{63}\right)  $ follows directly from
$\left(  \ref{75}\right)  $. For smaller $m$ we will use the following three lemmas.

\begin{lemma}
\label{small_m} For any $m\leq4$ boundary defect with order-disorder b.c. the
strong disorder b.c. implies that $\partial X_{0}$ contains at least $\frac
{3}{4}N^{2}$ vertical bonds, provided $q$ is large enough. (Here $X_{0}$ is
the ordered component of the top ordered cube.)
\end{lemma}

\begin{proof}
Since $m\leq4$, the defect is at most $8$-cubes wide, and therefore contains
at most $36$ sites. If $\partial X_{0}$ had less than $3$ vertical bonds in
the column, then two sites $u,v$ of the bottom plaquette would belong to
$X_{0}$. Denoting by $M$ the size of the largest possible path in a graph with
$36$ sites, we would get $\left\vert \sigma_{u}-\sigma_{v}\right\vert \leq M$,
therefore contradicting the strong disorder b.c. for $q$ large enough.
\end{proof}

\begin{lemma}
\label{small_Db} Consider any boundary defect with any b.c. on the top. If
$D^{b}\leq\frac{N^{2}}{4}$, the strong disorder b.c. imply
\[
\sum_{i\notin J}\left\vert \partial X_{i}\right\vert +\left\vert \partial
^{2}X_{i}\right\vert \geq3N^{2}.
\]

\end{lemma}

\begin{proof}
We start with the case $D^{b}=0.$ Since the b.c. are strongly disordered, all
horizontal bonds of the first layer have to be disordered as well, and each of
them belong to $\partial^{2}X_{i}$ for some $i\notin J$, so their contribution
to the sum above is $4N^{2}$.

In the case $D^{b}=N^{2}/4$, the strong disordered b.c. implies that three or
four horizontal bonds in the first layer are disordered. If we have 4 such
disordered bonds, they all belong to some $\partial X_{i},$ and two of them
actually belong to some $\partial^{2}X_{i};$ if we have only three such bonds,
they all belong to some $\partial^{2}X_{i}.$ In any case, they contribute
$3N^{2}$ to the sum above.
\end{proof}

\begin{lemma}
\label{dis_bond} For any boundary defect with order--disorder b.c. with a
horizontal disordered bond at the level $z=1$, the contribution of horizontal
bonds to
\[
\sum_{i\notin J}\left\vert \partial X_{i}\right\vert +\left\vert \partial
^{2}X_{i}\right\vert ,
\]
is at least $N^{2}$.
\end{lemma}

\begin{proof}
If all four bonds of the horizontal plaquette $P$ at $z=1$ are disordered,
there has to be a vertical ordered bond touching the boundary (because the
first cube is frustrated). It touches two horizontal bonds of $P$; all their
reflections contribute $N^{2}$ to the sum. If the plaquette $P$ has two or
three disordered bonds, at least two of them belong to the boundary of some
$X_{j}$. Finally, if $P$ has only one disordered bond, then it belongs to
$\partial^{2}X_{j}$ for some $j,$ and so contributes twice to the sum above.
\end{proof}

If $D^{b}\leq\frac{N^{2}}{4}$, we can apply Lemma \ref{small_Db} and $\left(
\ref{78}\right)  $ to get $\left(  \ref{63}\right)  $. If $D^{b}=\frac{N^{2}%
}{2}$, the strong disorder b.c. prevent the horizontal plaquette at $z=1$ from
being completely ordered, so we can apply Lemma \ref{dis_bond} together with
Lemma \ref{small_m}, getting $\left(  \ref{63}\right)  $.

If $D^{b}\geq\frac{3N^{2}}{4}$ and $m \geq3$, we apply $\left(  \ref{75}%
\right)  $ to get the relation $\left(  \ref{63}\right)  $.

In the remaining case $D^{b}\geq\frac{3N^{2}}{4}$ and $m=2$ we know, that the
blob corresponding to the defect had at least 2 plaquettes, because it would
be e-problematic otherwise. If this extra (disordered!) plaquette is
horizontal, then the second cube is pure disordered; else it is vertical. In
any case the horizontal plaquette at $z=1$ cannot be completely ordered. Thus
we can apply Lemma \ref{dis_bond} and $\left(  \ref{78}\right)  $ to get
$\left(  \ref{63}\right)  $.

$\blacksquare$

\subsection{Disorder--disorder: Proof of $\left(  \ref{70bis}\right)  $}

Now we prove the relation
\begin{equation}
2D-\frac{6\left(  L-\frac{1}{4}\right)  }{L}(K+Q)\geq\frac{7N^{2}}{2}+\alpha
mN^{2}. \label{70bis}%
\end{equation}
In fact, for most defects we will prove the stronger statement $\left(
\ref{71}\right)  :2D-6(K+Q)\geq\frac{7N^{2}}{2}+\alpha mN^{2}.\ $We start with
the identity
\begin{equation}
2D-6K=D^{b}+N^{2}+\sum_{i}\left\vert \partial X_{i}\right\vert +\sum
_{i}\left\vert \partial^{2}X_{i}\right\vert , \label{cr2}%
\end{equation}

\noindent where $D^{b}$ is the number of vertical disordered bonds attached to
the bottom boundary of $F^{N}$. (The term $N^{2}$ equals to the number of
vertical disordered bonds attached to the top boundary.) From this we deduce,
as above, that
\begin{equation}
2D-6K-6Q\geq D^{b}+N^{2}+\frac{mN^{2}}{2}+2Q. \label{92}%
\end{equation}
The desired estimate is directly derived from this for $m\geq6$. We now deal
with the case $m\leq5$.

The case $m=1$ is completely explicit. We have $L=1$, $D=3N^{2}+D^{b}$, $Q=0$,
$K=D^{b}$ and $D^{b}\leq\frac{3}{4}N^{2}$ (because the first cube is
frustrated), so that $2D-6\frac{L-1/4}{L}K=6N^{2}+2D^{b}-\frac{9}{2}%
D^{b}=6N^{2}-\frac{5}{2}D^{b}\geq4N^{2}+\frac{1}{8}N^{2}$, and thus we assume
$m\geq2$.

Also, if $K=Q=0$, the simple fact that $D\geq3N^{2}$ is enough to get
(\ref{71}), so we assume $K+Q\geq\frac{N^{2}}{4}.$

\begin{lemma}
\label{horiz_dis} For any boundary defect with $m\geq2$ and all horizontal
bonds disordered,
\[
D^{b}+ \sum\left\vert \partial X_{j}\right\vert \geq3N^{2} +6Q.
\]

\end{lemma}

\begin{proof}
Our assumption implies that all ordered components are vertical segments ;
since the first cube must be frustrated, $D^{b}\leq\frac{3}{4}N^{2}$. For
$j\notin J$, $X_{j}$ starts from the boundary and $\sum\left\vert \partial
X_{j}\right\vert \geq10Q+\sum_{j\notin J}\left(  4\left\vert X_{j}\right\vert
+1\right)  \geq10Q+5\left\vert J^{c}\right\vert $. Since $\left\vert
J^{c}\right\vert =N^{2}-D^{b}$, we have $D^{b}+\sum\left\vert \partial
X_{j}\right\vert \geq10Q+5N^{2}-4D^{b}\geq10Q+2N^{2}$. The lemma is then
proved if $Q\geq\frac{1}{4}N^{2}$, so we assume $Q=0$.

We now pick one ordered bond in the second frustrated cube, which is vertical
by assumption. Since $Q=0$, the corresponding ordered component $X_{j}$
satisfies $\left\vert X_{j}\right\vert \geq2$. After reflections, there are
$\frac{N^{2}}{4}$ such segments, and $\frac{3N^{2}}{4}-D^{b}$ other segments.
Then $D^{b}+\sum\left\vert \partial X_{j}\right\vert \geq D^{b}+\frac{9}%
{4}N^{2}+5\left(  \frac{3N^{2}}{4}-D^{b}\right)  =6N^{2}-4D^{b}$ and the lemma
follows from $D^{b}\leq\frac{3}{4}N^{2}$.
\end{proof}

\begin{lemma}
\label{horiz_ord} For any boundary defect with disorder-disorder b.c., and for
any ordered horizontal bond $e$,
\[
\left\vert \partial X_{e}\right\vert +\left\vert \partial^{2}X_{e}\right\vert
\geq N^{2}+\frac{3}{4}N^{2},
\]
where $X_{e}$ denotes the ordered component containing $e$.
\end{lemma}

\begin{proof}
We denote by $P$ the horizontal plaquette containing $e$. If $P$ is completely
ordered, the strong disorder b.c. force $\partial X_{e}\geq N^{2}+\frac{3}%
{4}N^{2}$ (compare with Lemma \ref{small_m}). If $P$ is not completely
ordered, either two horizontal bonds of $P$ belong to $\partial X_{e}$ or one
of them is in $\partial^{2}X_{e}$ ; moreover, due to the strong disorder b.c.
at least $3$ vertical bonds of the column belong to the boundary of $X_{e}$.
\end{proof}

If no horizontal bond is ordered, we can combine $\left(  \ref{cr2}\right)  $
with Lemma \ref{horiz_dis} to get $2D-6(K+Q)\geq4N^{2}$.

Otherwise, we can find a bond $e$ to which we apply Lemma \ref{horiz_ord}, to
get $2D-6(K+Q)\geq2N^{2}+D^{b}+\frac{3}{4}N^{2}$. If $D^{b}\geq\frac{1}%
{2}N^{2}$, this shows that $2D-6(K+Q)\geq3N^{2}+\frac{1}{4}N^{2}$.
Since $2D-6(K+Q)$ is an integer multiple of $\frac{1}{2}N^{2}$, we
actually have $2D-6(K+Q)\geq\frac{7}{2}N^{2}$. (\ref{70bis}) now
follows from $L\leq2m=10$ and $K+Q\geq\frac{N^{2}}{4}$.

If $D^{b}\leq\frac{1}{4}N^{2}$ we apply Lemma \ref{small_Db} and (\ref{cr2})
to get $2D-6(K+Q)\geq4N^{2}$. $\blacksquare$

\section{Proof of the Main Theorem \ref{T}}

In this section, we derive our main results from the Peierls estimate.

\noindent We start with the question of the interface uniqueness.

\begin{lemma}
For any $b>1$ there exists a $q_{0}<\infty$ such that the following holds: For
any $q \geq q_{0}$ and any sequence $L_{N}\leq b^{N^{2}}$ the probability
$\mu_{N,L_{N}}^{\beta,q}\left(  \text{discn}\right)  $ of the event that the
interface $B$ is disconnected, vanishes as $N\rightarrow\infty.$
\end{lemma}

\begin{proof}
Let $\mathbf{0\in T}$ be the origin. Denote by $l\left(  B\right)  $ the
quantity $\min\left\{  z:\left(  \mathbf{0},z\right)  \in B\right\}  ;$ it is
the height of the interface $B$ at the origin. Let $m\left(  B\right)  $ be
the number of frustrated cubes having at least one plaquette in common with
$B.$

Let $B$ be disconnected. Then it has at least three connected components,
which are interfaces themselves. Let $B_{1},B_{2},B_{3}$ be the first three of
them. Clearly,
\begin{align*}
&  \mathbf{\Pr}\left(  B\text{ is disconnected}\right) \\
&  =\sum_{l_{1}<l_{2}<l_{3}}\mathbf{\Pr}\left(  B_{1},B_{2},B_{3}:l\left(
B_{1}\right)  =l_{1},l\left(  B_{2}\right)  =l_{2},l\left(  B_{3}\right)
=l_{3}\right)  .
\end{align*}
Applying the Proposition \ref{P} we have
\begin{align*}
&  \mathbf{\Pr}\left(  B_{1},B_{2},B_{3}:l\left(  B_{1}\right)  =l_{1}
,l\left(  B_{2}\right)  =l_{2},l\left(  B_{3}\right)  =l_{3}\right) \\
&  \leq a^{m\left(  B_{1}\right)  +m\left(  B_{2}\right)  +m\left(
B_{3}\right)  -N^{2}}.
\end{align*}
Note that for any $B$ the number $m\left(  B\right)  \geq N^{2},$ and the
number of interfaces $B$ with $m\left(  B\right)  =m$ and with $l\left(
B\right)  $ fixed is at most $C^{m}$ for some $C.$ Therefore
\begin{align*}
&  \mathbf{\Pr}\left(  B\text{ has at least three components}\right) \\
&  \leq L_{N}^{3}\sum_{m\geq3N^{2}}\left(  Ca^{\frac{2}{3}}\right)
^{m}=const\cdot\left(  L\left(  Ca^{\frac{2}{3}}\right)  ^{N^{2}}\right)
^{3},
\end{align*}
which goes to zero as $N\rightarrow\infty$ once $L_{N}<b^{N^{2}}$ with
$b<\left(  Ca^{\frac{2}{3}}\right)  ^{-1}.$
\end{proof}

In what follows we will treat only connected interfaces. We will now show that
typically the interface does not have a wall which winds around the torus. The
reason is that such walls contain so many plaquettes that they appear very
seldom, as estimates from previous sections will show. We say that a wall
$\gamma$ is winding if the projection $\Pi\left(  \gamma\right)  $ contains a
non-trivial loop of the torus. In that case $\gamma$ contains at least $N$
plaquettes, so
\[
\mu_{N,L_{N}}^{\beta,q}\left(  \text{ there is a winding wall }\right)
\leq2N^{2}L_{N}\sum_{l\geq N}C^{l}a^{l/2},
\]
which goes to zero as $N\rightarrow\infty,$ once $L_{N}<b^{N}$ with $b<\left(
Ca^{1/2}\right)  ^{-1}.$

Let $M\in\mathbb{T}_{N}$ be a point in the 2D torus, and $\gamma$ be a wall of
some interface in the 3D box $\Lambda_{N,L}.$ Denote by $\tilde{\gamma}$ the
projection $\Pi\left(  \gamma\right)  .$ We will say that $\gamma$ surrounds
$M,$ iff $M\in\tilde{\gamma}\cup\mathrm{Int}\left(  \tilde{\gamma}\right)  .$
Evidently, the rigidity property of the interface that we want to prove, would
follow from the

\begin{proposition}
\label{PP}%
\begin{equation}
\mu_{N,L}^{\beta,q}\left(  M\text{ is surrounded by some wall}\right)  \leq
c\left(  q\right)  , \label{005}%
\end{equation}
with $c\left(  q\right)  \rightarrow0$ as $q\rightarrow\infty.$\medskip
\end{proposition}

\textbf{Remark. }The long-range claim of our main Theorem \ref{T}
also follows from the Proposition \ref{PP}. Indeed, if the heights
$h\left(  M^{\prime }\right)  \neq h\left(  M^{\prime\prime}\right)
$ or one of them is infinite, then at least one of the points
$M^{\prime},M^{\prime\prime}$ is surrounded by a wall.\medskip

\textbf{Proof. }From the Peierls estimate we know that the probability of the
presence of an interface wall $\gamma$ satisfies
\[
\mu_{N,L}^{\beta,q}\left(  \gamma\right)  \leq a^{w\left(  \gamma\right)  }.
\]
However, we need evidently the estimate on the probability of the larger event
$\gamma^{\ast}=\cup_{\tau}\gamma^{\tau},$ where $\gamma^{\tau}$ is the wall
obtained from $\gamma$ by a vertical shift along the vector $\left(
0,0,\tau\right)  .$ Since there are about $L$ values of $\tau$ for which
$\gamma^{\tau}\subset\Lambda_{N,L},$ the estimate we have thus far is
\[
\mu_{N,L}^{\beta,q}\left(  \gamma^{\ast}\right)  \leq La^{w\left(
\gamma\right)  },
\]
and since $L$ is diverging with $N,$ the above estimate seems to be not enough
for our purposes.

Yet, we know more about our measure $\mu_{N,L}^{\beta,q}.$ Namely, we know
also that if $\Gamma$ is a collection $\left(  \gamma_{1},...,\gamma
_{k}\right)  $ of the walls belonging to the same interface, then
\[
\mu_{N,L}^{\beta,q}\left(  \Gamma\right)  \leq a^{w\left(  \Gamma\right)  },
\]
with $w\left(  \Gamma\right)  =\sum w\left(  \gamma_{i}\right)  .$ Therefore
if $\Gamma^{Ext}=\left(  \gamma_{1},...,\gamma_{k}\right)  $ is a collection
of \emph{exterior }walls in some interface, and $\Gamma^{Ext\ast}=\left(
\gamma_{1}^{\ast},...,\gamma_{k}^{\ast}\right)  $ is the event $\cup_{\tau
_{1},...,\tau_{k}}\left(  \gamma_{1}^{\tau_{1}},...,\gamma_{k}^{\tau_{k}%
}\right)  $ of observing the collection $\left(  \gamma_{1}^{\tau_{1}%
},...,\gamma_{k}^{\tau_{k}}\right)  $ to be exterior walls of an interface,
then necessarily $\tau_{1}=...=\tau_{k},$ and so
\[
\mu_{N,L}^{\beta,q}\left(  \Gamma^{Ext\ast}\right)  \leq La^{w\left(
\Gamma\right)  }.
\]
This estimate is helpful to eliminate long walls, but it is useless for
dealing with collections $\Gamma$ of walls of finite total length $l$, when
$L$ is large. Note however that such a collection $\Gamma$ can surround only a
finite total area $\leq l^{2}$. Since our measure $\mu_{N,L}^{\beta,q}$ is
translation invariant, the probability to see $\Gamma$ at any given location
can be estimated by $\frac{l^{2}}{N^{2}}.$ In what follows we will make these
heuristic arguments rigorous.

We start with the following simplified model, which contains all the essential
features of our problem. Let $\mathbb{T}_{R}$ be a $R\times R$ discreet torus,
and let $\xi_{t}=0,1$ be a random field indexed by $t\in\mathbb{T}_{R}.$ Let
$\mu$ be the distribution of the field $\xi.$

\begin{lemma}
Suppose that

\begin{itemize}
\item $\mu$ is translation-invariant,

\item there exist a value $K$ such that for all $k\geq K$
\begin{equation}
\mu\left(  \xi_{t_{1}}=\xi_{t_{2}}=...=\xi_{t_{k}}=1\right)  \leq\alpha^{k},
\label{002}%
\end{equation}
for small enough $\alpha.$ Then there exists a function $R\left(
\alpha\right)  ,$ such that for any $t\in\mathbb{T}_{R}$
\[
\mu\left(  \xi_{t}=1\right)  \leq3\alpha,
\]
provided $R\geq\sqrt{\frac{K}{3\alpha}}+R\left(  \alpha\right)  .$
\end{itemize}
\end{lemma}

\begin{proof}
Let us show first that for any $k$
\begin{equation}
\mu\left(  \xi_{0}=1,\sum_{t\in\mathbb{T}_{R}}\xi_{t}=k\right)  =\frac
{k}{R^{2}}\mu\left(  \sum_{t\in\mathbb{T}_{R}}\xi_{t}=k\right)  . \label{001}%
\end{equation}
To see this let $X\subset\mathbb{T}_{R}$ be any subset with

$\left|  X\right|  =k,$ and let $\tilde{X}$ be the event that the support
$\mathrm{Supp}\left(  \xi\right)  $ coincides with some shift $X+t$ of $X,$
$t\in\mathbb{T}_{R}.$ Note that for any $X$
\[
\mu\left(  \xi_{0}=1\Bigm|\tilde{X}\right)  =\frac{k}{R^{2}}.
\]
This is immediate if the set $X$ is not periodic, i.e. if all the shifts $X+t$
are different subsets. In case $X$ is periodic we have to consider the
sublattice $\mathcal{L}_{X}\subset\mathbb{T}_{R}$ of its periods and its
fundamental parallelogram $\mathcal{P}_{X}\subset\mathbb{T}_{R}.$ By the same
reasoning $\mu\left(  \xi_{0}=1\Bigm|\tilde{X}\right)  =\frac{\left|
X\cap\mathcal{P}_{X}\right|  }{\left|  \mathcal{P}_{X}\right|  },$ while
evidently $\frac{\left|  X\cap\mathcal{P}_{X}\right|  }{\left|  \mathcal{P}%
_{X}\right|  }=\frac{k}{R^{2}}.$ Finally, $\mu\left(  \xi_{0}=1,\sum
_{t\in\mathbb{T}_{R}}\xi_{t}=k\right)  =\sum_{\tilde{X}}\mu\left(  \xi
_{0}=1,\tilde{X}\right)  =\frac{k}{R^{2}}\sum_{\tilde{X}}\mu\left(  \tilde
{X}\right)  =\frac{k}{R^{2}}\mu\left(  \sum_{t\in\mathbb{T}_{R}}\xi
_{t}=k\right)  .$

We now prove our lemma. From $\left(  \ref{001}\right)  $ we know that for all
$M\geq1$,
\begin{equation}
\mu\left(  \xi_{0}=1,1\leq\sum_{t\in\mathbb{T}_{R}}\xi_{t}\leq M\right)
=\sum_{k=1}^{M}\frac{k}{R^{2}}\mu\left(  \sum_{t\in\mathbb{T}_{R}}\xi
_{t}=k\right)  \leq\frac{M}{R^{2}}. \label{003}%
\end{equation}
In the region $\sum_{t\in\mathbb{T}_{R}}\xi_{t}\geq M$ we would like to use
the \textquotedblleft Peierls estimate\textquotedblright\ (\ref{002}).

Our choice will be $M=bR^{2}$ with some $b\geq2\alpha.$ Then we have
\begin{equation}
\mu\left(  \xi_{0}=1,\sum_{t\in\mathbb{T}_{R}}\xi_{t}\geq M\right)  \leq
\binom{R^{2}}{M}\alpha^{M}. \label{004}%
\end{equation}
Once $R$ satisfies $bR^{2}>K,$ we can use both $\left(  \ref{003}\right)  $
and $\left(  \ref{004}\right)  $ to conclude that
\[
\mu\left(  \xi_{0}=1\right)  \leq b+\binom{R^{2}}{bR^{2}}\alpha^{bR^{2}}.
\]
Finally, introducing $c=1-b,$ we have by Stirling, that
\begin{align*}
\binom{R^{2}}{bR^{2}}\alpha^{bR^{2}}  &  \sim\frac{R^{2R^{2}}}{\sqrt{2\pi
bcR^{2}}\left(  bR^{2}\right)  ^{bR^{2}}\left(  cR^{2}\right)  ^{cR^{2}}%
}\alpha^{bR^{2}}\\
&  =\frac{1}{\sqrt{2\pi bcR^{2}}}\left(  \frac{\alpha^{b}}{b^{b}c^{c}}\right)
^{R^{2}}.
\end{align*}
A straightforward check shows that for $b=3\alpha$ the ratio $\frac{\alpha
^{b}}{b^{b}c^{c}}<1,$ so $\binom{R^{2}}{bR^{2}}\alpha^{bR^{2}}\rightarrow0$ as
$R\rightarrow\infty,$ which concludes the proof.
\end{proof}

Returning to the proof of Proposition \ref{PP}, we take two large numbers, $R$
and $Q,$ to be chosen later, and we consider the box $\Lambda_{N,L}$ with
$N=QR$. Let $O$ be the origin, $O\in\mathbb{T}_{N}.$ The probability that $O$
is surrounded by a wall with the weight $w\geq Q$ satisfies
\begin{equation}
\mu_{N,L}^{\beta,q}\left(  O\text{ is }Q\text{-surrounded}\right)  \leq La^{Q}
\label{006}%
\end{equation}
(modulo unimportant constant). So once $Q\gg\ln L,$ this probability is small.
We are left with the event that $O$ is surrounded by a wall with the weight
$w<Q.$ To estimate its probability we will use the above Lemma. Let us
consider the torus sublattice $\mathbb{T}_{R}\subset\mathbb{T}_{N}.$ For every
point $t\in\mathbb{T}_{R}$ we define the random variable $\xi_{t}$ by
\[
\xi_{t}=\left\{
\begin{array}
[c]{cc}%
1 & \text{ if }t\text{ is inside some }\tilde{\gamma}\text{ with }w\left(
\gamma\right)  <Q,\\
0 & \text{ otherwise.}%
\end{array}
\right.
\]
Evidently, the field $\left\{  \xi_{t},t\in\mathbb{T}_{R}\right\}  $ is
translation-invariant. Let us estimate the probability of the event
\[
\xi^{T}=\left\{  \xi_{t}=1\text{ for all }t\in T\subset\mathbb{T}_{R}\right\}
.
\]
As was explained above, our Peierls estimate gives
\[
\mu_{N,L}^{\beta,q}\left(  \xi^{T}\right)  \leq L\sum_{\gamma_{1}%
,...,\gamma_{\left|  T\right|  }}a^{w\left(  \gamma_{1}\right)  +...+w\left(
\gamma_{\left|  T\right|  }\right)  },
\]
where the summation goes over all collections $\left\{  \gamma_{1}%
,...,\gamma_{\left|  T\right|  }\right\}  $ of exterior walls with base at the
given level -- say, $L/2$ -- such that every $\gamma_{i}$ surrounds precisely
one point of $T.$ Since always $w\left(  \gamma\right)  \geq4,$
\[
\mu_{N,L}^{\beta,q}\left(  \xi^{T}\right)  \leq La^{4\left|  T\right|  }.
\]
Therefore the condition $\left(  \ref{002}\right)  $ holds with $\alpha=a^{2}$
and $K=\frac{1}{2}\ln L\ln\frac{1}{a}.$ Hence for
\begin{equation}
R\geq\frac{\sqrt{\ln L\ln\frac{1}{a}}}{a} \label{007}%
\end{equation}
we have $\mu_{N,L}^{\beta,q}\left(  \xi_{O}=1\right)  \leq3a^{2}.$ On the
other hand, our bound $La^{Q}$ from $\left(  \ref{006}\right)  $ satisfies
$La^{Q}<a^{2}$ once
\begin{equation}
\left(  Q-2\right)  \ln\frac{1}{a}>\ln L. \label{008}%
\end{equation}
If we take $Q=R^{2}$, then both inequalities $\left(  \ref{007}\right)  $ and
$\left(  \ref{008}\right)  $ will be satisfied, provided $\ln L<N^{2/3}.$ So
under this condition we can conclude that $\left(  \ref{005}\right)  $ is
satisfied with $c\left(  q\right)  =4a^{2}.$

$\blacksquare$

\section{Conclusions}

In this work we have developed a version of the Reflection Positivity method
suitable for the investigation of the rigidity property of the interfaces
between coexisting phases of certain 3D systems. It is applicable to various
known models, such as the Ising, Potts or FK models. However, the main
advantage of the method is that it works also for models with non-trivial
structure of the ground states, which can not be treated by the PS theory, one
example being the clock version of the \textquotedblleft very non-linear
$\sigma$-model\textquotedblright.

We hope to be able to extend our methods to systems with continuous symmetry.

\textbf{ Acknowledgement.} \textit{In the course of this work we have
benefitted by discussions with many colleagues, including M. Biskup, L.
Chayes, A. van Enter, K. Khanin, S. Miracle-Sole, Ch. Pfister, S. Pirogov, to
whom we express our gratitude. The generous support of GREFI MEFI is
gratefully acknowledged. S.S. also acknowledges the support of the Grant }
\textit{05-01-00449 of RFFR.}


\begin{thebibliography}{9999}                                                                                             %


\bibitem[A]{A}Michael Aizenman: On the slow decay of O(2) correlations in the
absence of topological excitations: Remark on the Patrascioiu-Seiler model,
Journal of Statistical Physics, v.77, pp. 351-359, 1994.

\bibitem[CK]{CK}J. Cerny, R. Kotecky: Interfaces for random cluster models.
Journal of statistical physics 111, 73-106, 2003.

\bibitem[D72]{D72}R. L. Dobrushin. Gibbs state, describing the coexistence of
phases in the three-dimensional Ising model. Th. Prob. and its Appl., 17,
582-600, 1972.

\bibitem[DS]{DS}R.L. Dobrushin and S. Shlosman: \textit{Phases corresponding
to the local minima of the energy}, Selecta Math. Soviet. 1 (1981), no. 4, 317--338

\bibitem[D]{D}A. Dold. Lectures on Algebraic Topology. Springer, 1995.

\bibitem[FILS]{FILS}J. Fr\"{o}hlich, R. Israel, E. Lieb, B. Simon. Phase
transitions and reflection positivity I. Comm. Math. Phys., 62, pp.1-34, 1978.

\bibitem[FL]{FL}J\"{u}rg Fr\"{o}hlich and Elliott H. Lieb. Phase transitions
in anisotropic lattice spin systems. Source: Comm. Math. Phys. 60, no. 3
(1978), 233--267

\bibitem[FP]{FP}J\"{u}rg Fr\"{o}hlich and Charles-Edouard Pfister: Spin waves,
vortices, and the structure of equilibrium states in the classical XY model,
CMP, v. 89, pp. 303-327, 1983.

\bibitem[FSS]{FSS}J. Frohlich, B. Simon and T. Spencer, Infrared bounds, phase
transitions and continuous symmetry breaking, Commun. Math. Phys. 50, 79 (1976).

\bibitem[FS]{FS}J. Frohlich and T. Spencer: The Kosterlitz-Thouless transition
in two-dimensional Abelian spin systems and the Coulomb gas, Comm. Math.
Phys., 81, 1981, pp. 527-602.

\bibitem[G]{G}G. Grimmett, private communication.

\bibitem[GG]{GG}G. Gielis, G. Grimmett: Rigidity of the Interface in
Percolation and Random-Cluster Models, Journal of Statistical Physics, Volume
109, Numbers 1-2, 2002, pp 1 - 37.

\bibitem[HKZ]{HKZ}P. Holicky, R. Kotecky, and M. Zahradn\i k, Rigid interfaces
for lattice models at low temperatures, J. Statist. Phys. 50:755--812 (1988).

\bibitem[K]{K}Richard Kenyon: Dominos and the Gaussian free field, Ann. Prob.
29, no. 3 (2001), 1128-1137.

\bibitem[ES1]{ES1}van Enter, A. C. D. and Shlosman, S.: First-Order
Transitions for n-Vector Models in Two and More Dimensions: Rigorous Proof,
Phys. Rev. Lett., 89, 285702, 2002.

\bibitem[ES2]{ES2}van Enter, A. C. D. and Shlosman, S.: Provable First-Order
Transitions for Nonlinear Vector and Gauge Models with Continuous Symmetries,
Comm. Math. Phys., Volume 255, Number 1, 2005, pp 21 - 32.

\bibitem[S]{S}Senya Shlosman: \textit{The Method of Reflection Positivity in
the Mathematical Theory of First-Order Phase Transitions,} Russian Math.
Surveys, \textbf{41}:3, 83-134, 1986.

\bibitem[SV]{SV}Senya Shlosman and Yvon Vignaud: Rigidity of the interface
between low-energy and high-entropy phases. In preparation.

\bibitem[V]{V}Yvon Vignaud: Entropic repulsion and entropic attraction. In preparation.

\bibitem[V1]{V1}Yvon Vignaud: Rigidity of the interface for a continuous symmetry model in a slab. In preparation.

\end{thebibliography}
\end{document}